\documentclass[]{article}

\usepackage{amsmath}
\usepackage{amssymb}
\usepackage{graphicx} 
\usepackage{epsfig} 
\usepackage{subfig} 
\usepackage{verbatim} 
\usepackage{array} 

\title{Evolutionary Processes in Finite Populations}

\author{Dirk M. Lorenz$^1$, Jeong-Man Park$^{1,2}$ and Michael W. Deem$^1$\\
$^1$Department of Physics, Rice University, Houston, TX\\
$^2$Department of Physics, The Catholic University of Korea, Bucheon, Korea
}

\begin{document}
\renewcommand{\_}[1]{{}_{\mathrm{#1}}} 
\newcommand{\phys}[1]{{\textsc{#1}}} 
\newcommand{\bea}{\begin{eqnarray*}} 
\newcommand{\eea}{\end{eqnarray*}} 
\newcommand{\ansbox}[1]{{\framebox{$\displaystyle#1$}}} 
\newcommand{\norm}[1]{\left|#1\right|} 
\newcommand{\normTwo}[1]{\left\|\,#1\,\right\|} 
\newcommand{\e}[1]{{\vec e}\_{#1}} 
\newcommand{\blankline}{\\ \par \noindent} 
\newcommand{\grad}{\vec \nabla} 
\newcommand{\pwrt}[2]{\frac{\partial #1}{\partial #2}}
\newcommand{\curlmat}[3]{
    \begin{vmatrix}
        \e 1 & \e 2 & \e 3 \\ \\
        \displaystyle\pwrt{}{x_1}
            & \displaystyle\pwrt{}{x_2}
            & \displaystyle\pwrt{}{x_3} \\ \\
        \displaystyle#1 & \displaystyle#2 & \displaystyle#3
    \end{vmatrix}} 
\newcommand{\Res}[2]{\operatorname{Res}\left[#1, #2\right]}
\newcommand{\Resf}[1]{\operatorname{Res}\left[f, #1\right]}
\newcommand{\Real}[1]{\operatorname{Re}\left(#1\right)}
\newcommand{\Imaginary}[1]{\operatorname{Im}\left(#1\right)}
\newcommand{\Span}[1]{\operatorname{Sp}\left\{#1\right\}}
\newcommand{\expval}[1]{\left<#1\right>} 
\newcommand{\centerthis}[1]{\vphantom{}\hfill #1 \hfill\vphantom{}} 
\newcommand{\BraP}[1]{\left(#1\right|}
\newcommand{\Bra}[1]{\left<#1\right|}
\newcommand{\KetP}[1]{\left|#1\right)}
\newcommand{\Ket}[1]{\left|#1\right>}
\newcommand{\BraKet}[2]{\left<#1\vphantom{#2}\right.\left|#2\vphantom{#1}\right>}
\newcommand{\BraKetP}[2]{\left<#1\vphantom{#2}\right.\left|#2\vphantom{#1}\right)}
\newcommand{\BraPKetP}[2]{\left(#1\vphantom{#2}\right.\left|#2\vphantom{#1}\right)}
\newcommand{\Diag}[1]{\operatorname{diag}\left(#1\right)}

\makeatletter
\def\arrowfill{$\m@th\mathord=\mkern-7mu
  \cleaders\hbox{$\!\mathord{=}\!$}\hfill
  \mkern-7mu\mathord\Rightarrow$}
\makeatother
\newcommand{\strRightarrow}{\hbox{\arrowfill}}

\renewcommand{\vec}{\boldsymbol}

\maketitle

\begin{abstract}
We  consider the evolution of large but finite populations on arbitrary
fitness landscapes.  We describe the evolutionary process by a Markov, Moran
process.  We show that to $\mathcal O(1/N)$, the time-averaged fitness is lower
for the finite population than it is for the infinite population.  We also show that fluctuations
in the number of individuals for a given genotype  can be
proportional to a power of the inverse of the mutation rate.  Finally, we
show that the probability for the system to take a given path through
the fitness landscape can be non-monotonic in system size.
\end{abstract}

\section{Introduction}
Natural populations are characterized by finite sizes. For this reason, it is impossible for biology to sample the entire space of all possible genotypes. Even the number of possible sequences with high fitness is typically much larger than the population size in naturally occurring populations. Effects due to finite population size are particularly pronounced in asexual populations.  For example, the reduction of fitness in a finite population without back mutation is termed Muller's ratchet \cite{Muller}, and the decreased speed of evolution in a finite population without recombination is termed the the Hill-Robertson effect \cite{Hill}.

The relative influence of different evolutionary forces changes between small and large populations. While stochastic effects such as genetic drift act more strongly on small populations, natural selection acts more effectively on large populations. Many results in classical population genetics have focused on the limiting cases of small or infinite populations. In sufficiently small populations, beneficial mutations occur but rarely survive long enough to become established in the population. Those  mutations that survive, however, can spread through a small population, reaching fixation, before another beneficial mutation arises. This regime is referred to as successional-mutations regime \cite{Desai2007,Sella} and is fairly well-understood.  This theory has been useful, for example, to understand evolution of transcription factor binding sites \cite{Berg}. As the population size increases, beneficial mutations arise more frequently. Fixation of individual mutations does not occur before the arrival of another beneficial mutation. In asexual populations this leads to competition between descendants of each of the mutations --- an effect referred to as clonal interference \cite{Gerrish1998}. As the population becomes even larger, ultimately stochastic effects become negligible, and the time-evolution of the evolving population can be described by a set of ordinary differential equations. This regime has been studied extensively in quasispecies theory, albeit often only for simple fitness functions.

Here we investigate the regime between clonal interference and quasispecies theory. We seek to predict the evolutionary dynamics followed by a large yet finite population and how this dynamics differs from that of an infinite population. The study of finite-population effects requires a stochastic description based on a master equation \cite{Alves1998}. We make no assumption about the fitness landscape upon which the population evolves. We show that, averaged over time, the average fitness of a large finite population is lower than that of a population of infinite size. In other words, for large asexual populations evolving on a fixed fitness landscape, an increase in population size is accompanied by an increase in the average fitness. Furthermore, small mutation rates lead to high fluctuations and correlations. In particular, for small mutation rates, fluctuations and correlations in the number of individuals for a given genotype are inversely proportional to a power of the mutation rate. These large correlations enhance finite population effects and make the convergence to infinite-population behavior occur only for extremely large populations.

This article is organized as follows. We describe the stochastic process underlying our studies in section \ref{sec:stochasticProcess}.  We explain how this dynamic process can be written as a field theory.
We derive analytic results for the infinite population evolution from this field theory.
We describe finite population effects in section \ref{sec:finitePopShift}. We introduce the fitness landscape that we use to illustrate our results in section \ref{sec:landscape}.  In section \ref{sec:fluctuations} we investigate fluctuations in this random process. We verify our analytic results using stochastic simulations in section \ref{sec:compareToSimulations}.  We conclude in section \ref{sec:discussion}.

\section{Stochastic Process Mapped to a Field Theory}
\label{sec:stochasticProcess}
Throughout this article, we use the Moran process to model evolution of a population \cite{Moran}.  The individuals in the population are identified by their genotype, a sequence of length $l$. In this continuous-time process a constant population size, $N$, is maintained by simultaneous replication and death. The individual to be replicated is chosen randomly from the population with probability proportional to its microscopic fitness, while the individual to be killed is chosen randomly from the population with uniform probability. We further assume that replication and mutation are independent. Thus, there are two classes of events: mutation and replication. Mutation from genotype $i$ to genotype $j$ occurs at a rate of $\mu \Delta_{ij} N_i$, where $\mu$ is the mutation rate per locus, $N_i$ is the number of individuals with genotype $i$, and $\Delta_{ij}$ is equal to one if an individual can mutate from sequence $i$ to sequence $j$ with a single mutation and $\Delta_{ij}$ is equal to zero otherwise. This description allows for the incorporation of back-mutations which are often ignored in the literature. Note that the analytical results in this paper do not depend on this binary form of the matrix $\Delta$. Its elements can be arbitrary non-negative numbers as would be appropriate if back-mutation rates differed from forward mutation rates. Replication of genotype $i$ and simultaneous death of genotype $j$ occurs at a rate of $\frac{1}{N} r_i N_i N_j$, where $r_i$ is the replication rate of sequence $i$. The stochastic master equation for this process is
\begin{align}
    \label{eq:masterEquation}
    \notag
    \frac{\partial}{\partial t} P(\boldsymbol N; t)
        &= \mu \sum_{i,j} \Delta_{ij} \left[\left(N_i + 1\right) P(\boldsymbol N + \boldsymbol e_i - \boldsymbol e_j; t) - N_i P(\boldsymbol N; t)\right] \\
            &+ \frac{1}{N} \sum_{i} r_i \sum_{j \ne i} \left[\left(N_i - 1\right) \left(N_j + 1\right) P(\boldsymbol N - \boldsymbol e_i + \boldsymbol e_j; t)
                - N_i N_j P(\boldsymbol N; t)\right].
\end{align}
Here  $\boldsymbol N$ is a vector describing the state of the population by the number of individuals of each genoptype:
$(N_1, N_2, \ldots)$, and $ \boldsymbol e_i$ is a unit vector associated with genotype $i$.  Note $\sum_i N_i = N$.

\label{sec:mapToFieldTheory}
We obtain analytic expressions for the average occupation numbers and the fluctuations by mapping the stochastic process described in the previous section onto a field theory following \cite{Park2010}. To do this we introduce the state vector
\begin{align}
    \left|\psi(t)\right> = \sum_{\boldsymbol N} P(\boldsymbol N; t) \left|\boldsymbol N\right>
\end{align}
whose time evolution is governed by
\begin{align}
    \notag
    \frac{\partial}{\partial t} \left|\psi(t)\right>
        &= \sum_{\boldsymbol N}
            \left[
            \mu \sum_{i,j} \Delta_{ij} \left[\left(N_i + 1\right) P(\boldsymbol N + \boldsymbol e_i - \boldsymbol e_j; t) - N_i P(\boldsymbol N; t)\right] \right.\\
                &\qquad\left.+ \frac{1}{N} \sum_{i} r_i \sum_{j \ne i} \left[\left(N_i - 1\right) \left(N_j + 1\right)
                        P(\boldsymbol N - \boldsymbol e_i + \boldsymbol e_j; t)
                    - N_i N_j P(\boldsymbol N; t)\right]
            \right]
            \left|\boldsymbol N\right>.
\end{align}
By defining annihilation and creation operators
\begin{align}
    \hat a_i \left|\boldsymbol N\right> = N_i \left|\boldsymbol N - \boldsymbol e_i\right>,
    \qquad
    \hat a_i^\dagger \left|\boldsymbol N\right> = \left|\boldsymbol N + \boldsymbol e_i\right>
    \qquad
    \hat a_i \hat a_j^\dagger - \hat a_j^\dagger \hat a_i = \delta_{ij},
\end{align}
we can write the governing equation for the state vector as
\begin{align}
    \frac{\partial}{\partial t} \left|\psi(t)\right> = -\hat H \left|\psi(t)\right>,
\end{align}
where
\begin{align}
    - \hat H = \mu \sum_{i,j} \Delta_{ij} \left(\hat a_j^\dagger - \hat a_i^\dagger\right) \hat a_i
        + \frac{1}{N} \sum_{i,j} r_i \hat a_i^\dagger \left(\hat a_i^\dagger - \hat a_j^\dagger \right) \hat a_i \hat a_j.
\end{align}
This differential equation has the formal solution
\begin{align}
    \left|\psi(t)\right> = e^{- \hat H t} \left|\psi(0)\right>,
\end{align}
where $\left|\psi(0)\right> = \left|\boldsymbol N^0\right>$ is the initial distribution of individuals in the population. At time $T$, the average of an observable represented by the (normal-ordered) operator $F(\{\hat a_i^\dagger, \hat a_i\})$ can be obtained \cite{Peliti1985} by multiplying with the ``sum bra'' $\left<\cdot\right| = \left<0\right|(\prod_i e^{\hat a_i})$
\begin{align}
    \left<F\right>_T = \left<\cdot\right|F(\{\hat a_i^\dagger, \hat a_i\})\left|\psi(T)\right>
        = \left<\cdot\vphantom{N^0}\right|F(\{\hat a_i^\dagger, \hat a_i\})e^{- \hat H T} \left|\boldsymbol N^0\right>.
\end{align}
We introduce a Trotter factorization for the evolution operator $e^{- \hat H T}$, using a time interval $\epsilon \to 0$, in the basis of coherent states defined by $\hat a_i \left|\boldsymbol z\right> = z_i \left|\boldsymbol z\right>$ and obtain a path integral representation
\begin{align}
    \notag
    \left<\cdot\vphantom{N^0}\right|F(\{\hat a_i^\dagger, \hat a_i\})e^{- \hat H T} \left|\boldsymbol N^0\right>
        &= \left<\cdot\vphantom{N^0}\right|F(\{\hat a_i^\dagger, \hat a_i\})
               e^{-\epsilon \hat H} \cdot e^{-\epsilon \hat H} \cdot \, \cdots \, \cdot e^{-\epsilon \hat H} \left|\boldsymbol N^0\right> \\
        &= \int [Dz^*Dz] F(\{z(T/\epsilon)\}) e^{-S(\boldsymbol z, \boldsymbol z^*)}.
\end{align}
Here, the action in the exponent is, after the change of variables $z_i^* \equiv 1 + \bar z_i$,
\begin{align}
    \notag
    S(\boldsymbol z, \bar {\boldsymbol z})
        \label{eq:action}
        &= \sum_i\left[\sum_{k=0}^{T/\epsilon} \bar z_i(k) z_i(k) - \sum_{k=1}^{T/\epsilon} \bar z_i(k) z_i(k-1) - N_i(0) \ln \left(1 + \bar z_i(0)\right)\right] \\
           \notag
            &-  \mu \epsilon \sum_{k=1}^{T/\epsilon} \sum_{i,j} \left(\bar z_j(k) - \bar z_i(k)\right) z_i(k-1) \Delta_{i j}\\
            &- \frac{\epsilon}{N} \sum_{k=1}^{T/\epsilon} \sum_{i,j} r_i \left(1 + \bar z_i(k)\right) \left(\bar z_i(k) - \bar z_j(k)\right) z_i(k-1) z_j(k-1).
\end{align}

\label{sec:infPopEvolution}
The population dynamics in the limit as the population size, $N$, becomes infinite emerges as a saddle point in the action \cite{Park2010}. Setting $\left.\delta S / \delta z_i(t)\right|_c = 0$ leads to $\bar z_i^c(t) = 0$. From setting $\left.\delta S / \delta \bar z_i(t)\right|_c = 0$ we obtain $z_i^c(t) = N p_i(t)$ where $p_i(t)$ obeys the differential equation
\begin{align}
    \label{eq:diffEqP}
    \frac{dp_i}{dt} = \mu \sum_j \left(\Delta_{ji} p_j - \Delta_{ij} p_i\right) + r_i p_i - \left<r\right> p_i.
\end{align}
Here $\left<r\right> = \sum_j r_j p_j$ is the average fitness of the infinite population. This differential equation has the closed-form solution \cite{Thompson1974}
\begin{align}
    p_i(t) = \frac{\sum_j \left(e^{Yt}\right)_{ij} p_j(0)}{\sum_{a,j} \left(e^{Yt}\right)_{aj} p_j(0)},
\label{eq:closedFormP}
\end{align}
where the matrix $Y$ is defined by $Y_{ij} = \mu \Delta_{ji} - \mu \delta_{ij} \sum_k \Delta_{ik} + \delta_{ij} r_i$.

\section{Finite Population Shift to Probability Distribution}
\label{sec:finitePopShift}
We proceed to  quantify analytically how finite population effects alter the infinite population dynamics. To do so we expand the action about the saddle point and separate it into a Gaussian and a non-Gaussian part. Introducing $z_i(k) = {z_c}_i(k) + \delta z_i(k)$ and $\bar z_i(k) = \delta \bar z_i(k)$ in Eq.\ \ref{eq:action} we can write $S = S_0 + \Delta S$, where the reference action $S_0$ can be written as
\begin{align}
    S_0 = \frac12 \boldsymbol x^T \cdot \Pi_0^{-1} \cdot \boldsymbol x
\end{align}
where
\begin{align}
    \boldsymbol x^T = \left(\{\delta \bar {\boldsymbol z}(0), \delta \boldsymbol z(0)\},
        \{\delta \bar {\boldsymbol z}(1), \delta \boldsymbol z(1)\},
        \cdots,
        \{\delta \bar {\boldsymbol z}(T/\epsilon), \delta \boldsymbol z(T/\epsilon)\}\right).
\end{align}
Here,
\begin{align}
    \label{eq:Pi0Inv}
    \Pi_0^{-1}
        = \left(
        \begin{array}{c|c|c|c|c}
            \left({\Pi_0^{-1}}\right)_{00} & -\left({\Pi_0^{-1}}\right)_{01} & 0 & 0 & \cdots \vphantom{\text{\huge X}}\\[4pt] \hline
            -\left({\Pi_0^{-1}}\right)_{10} & \left({\Pi_0^{-1}}\right)_{11} & -\left({\Pi_0^{-1}}\right)_{12}
                & 0 & \cdots \vphantom{\text{\huge X}}\\[4pt] \hline
            0 & -\left({\Pi_0^{-1}}\right)_{21} & \left({\Pi_0^{-1}}\right)_{22} & -\left({\Pi_0^{-1}}\right)_{23}
                & \ddots \vphantom{\text{\huge X}}\\[4pt] \hline
            0 & 0 & -\left({\Pi_0^{-1}}\right)_{32} & \left({\Pi_0^{-1}}\right)_{33} & \ddots \vphantom{\text{\huge X}}\\[4pt] \hline
            \vdots & \vdots & \ddots & \ddots & \ddots \vphantom{\text{\huge X}}
        \end{array}
        \right)
\end{align}
with
\begin{align}
    \notag
    \left({\Pi_0^{-1}}\right)_{00} =
    \begin{pmatrix}
        \delta_{i j} N_i(0) & \delta_{i j} \\ \delta_{i j} & 0
    \end{pmatrix},
    &&
    \left({\Pi_0^{-1}}\right)_{kk} =
    \begin{pmatrix}
        -\epsilon (B)_{ij} & \delta_{i j} \\ \delta_{i j} & 0
    \end{pmatrix}, \\
    \left({\Pi_0^{-1}}\right)_{k,k-1} =
    \begin{pmatrix}
        0 & \delta_{i j} + \epsilon (A)_{ij} \\ 0 & 0
    \end{pmatrix},
    &&
    \left({\Pi_0^{-1}}\right)_{k-1,k} =
    \begin{pmatrix}
        0 & 0 \\ \delta_{i j} + \epsilon (A)^T_{ij} & 0
    \end{pmatrix}.
\end{align}
The matrices $A$ and $B$ are
\begin{align}
    \notag
    (A)_{ij} &= \mu \Delta_{j i} - \mu \delta_{i j} \left(\sum_m \Delta_{i m}\right)
            + \frac{1}{N} r_i {z_c}_i(k-1) + \delta_{i j} r_i \\
            & - \frac{1}{N} \delta_{i j} \left(\sum_m r_m {z_c}_m(k-1)\right)
            - \frac{1}{N} r_j {z_c}_i(k-1),
    \label{eq:matA}
\end{align}
and
\begin{align}
    (B)_{ij} =  2 \delta_{i j} r_i {z_c}_i(k-1) - \frac{1}{N} \left(r_i + r_j\right) {z_c}_i(k-1) {z_c}_j(k-1).
\end{align}
The non-Gaussian part of the action is given by
\begin{align}
    \notag
    \Delta S
        &=- \sum_i N_i(0) \left[ \ln \left(1 + \delta \bar z_i(0)\right)- \delta \bar z_i(0) + \frac12 (\delta \bar z_i(0))^2)\right] \\
        &- \frac{\epsilon}{N} \sum_{k=1}^{T/\epsilon} \sum_{i,j} \left[r_i \left(\delta \bar z_i(k) - \delta \bar z_j(k)\right)
            \delta z_i(k-1) \delta z_j(k-1) \right. \\
        \notag
        &\qquad+ r_i \delta \bar z_i(k) \left(\delta \bar z_i(k) - \delta \bar z_j(k)\right) {z_c}_i(k-1) \delta z_j(k-1)    \\
        \notag
        &\qquad+ r_i \delta \bar z_i(k) \left(\delta \bar z_i(k) - \delta \bar z_j(k)\right) \delta z_i(k-1) {z_c}_j(k-1) \\
       \notag
        &\qquad\left. + r_i \delta \bar z_i(k) \left(\delta \bar z_i(k) - \delta \bar z_j(k)\right) \delta z_i(k-1) \delta z_j(k-1) \right].
\end{align}

This formulation allows us to calculate averages using the Gaussian action and thermodynamic perturbation theory, which is equivalent to a cumulant expansion. The average occupation numbers are given by
\begin{align}
    \left<N_i\right>_T &= \left<\cdot\vphantom{N^0}\right|\hat a_i^\dagger \hat a_i e^{- \hat H T} \left|\boldsymbol N^0\right>
        = \left<\cdot\vphantom{N^0}\right|\hat a_i e^{- \hat H T} \left|\boldsymbol N^0\right>
        = \int [Dz^*Dz] z_i(T/\epsilon) e^{-S(\boldsymbol z, \bar {\boldsymbol z})} \\
        &= \int [Dz^*Dz] z_i(T/\epsilon) e^{-\Delta S} e^{-S_0}
        = \left<z_i(T/\epsilon) e^{-\Delta S}\right>_0 \\
        &= \left<z_i(T/\epsilon)\right>_0 -\left<z_i(T/\epsilon) \Delta S\right>_0  + \frac12 \left<z_i(T/\epsilon) (\Delta S)^2\right>_0 + \cdots \\
        &= N p_i(T) -\left<\delta z_i(T/\epsilon) \Delta S\right>_0  + \frac12 \left<\delta z_i(T/\epsilon) (\Delta S)^2\right>_0 + \cdots,
\end{align}
where the last step follows from $\left<(\Delta S)^n\right>_0 = 0$ $\forall n \in \mathbb Z, n \ge 1$. This procedure leads to an asymptotic expansion for the occupation numbers in powers of $1/N$. To first order, we obtain
\begin{align}
    \label{eq:firstCorrection}
    \frac{1}{N} \left<N_a\right>(T) \sim p_a(T)
        + \frac{1}{N^2} \int_0^T dt \, \sum_{i,j} {\Pi_0}^{z \bar z}_{a i}(T,t) {\Pi_0}^{z z}_{i j}(t,t) \left(r_i - r_j\right).
\end{align}
This expansion about infinite size is accurate when the correction term
on the right hand side of Eq.\ (\ref{eq:firstCorrection}) is much smaller than
$p_a(T)$.  Equation (\ref{36}) provides an estimate of the magnitude of the
correction for a common landscape with $k$ intermediate steps.
The second order term is given by Eq.\ \ref{eq:secondCorrection} in the appendix. We derive expressions for the matrices ${\Pi_0}^{z \bar z}_{a i}(T,t)$ and ${\Pi_0}^{z z}_{i j}(t,t)$ by inverting $\Pi_0^{-1}$ in Eq.\  \ref{eq:Pi0Inv}. In continuous time for $T > t$, they obey
\begin{align}
    \label{eq:diffEqPizzbar}
    \frac{\partial{\Pi_0}^{z \bar z}(T,t)}{\partial T} = A(T) {\Pi_0}^{z \bar z}(T,t),
\end{align}
with
\begin{align}
    \label{eq:initCondPizzbar}
    {\Pi_0}^{z \bar z}(t,t) = I
\end{align}
and
\begin{align}
    \label{eq:diffEqPizz}
    \frac{d{\Pi_0}^{z z}(t,t)}{dt} = B(t) + A(t) {\Pi_0}^{z z}(t,t) + {\Pi_0}^{z z}(t,t) A^T(t),
\end{align}
with
\begin{align}
    \label{eq:initCondPizz}
    {\Pi_0}^{z z}_{i j}(0,0) = -\delta_{i j} N_i(0).
\end{align}

Using the expression for the first-order shift to the occupation numbers due to finite population effects, we calculate the finite population shift in the average fitness of the population. The average fitness correction is
\begin{align}
    \label{eq:avgFitnessCorrection}
    \left<\delta r(T)\right> &= \frac{1}{N^2} \int_0^T dt \, \sum_{i,j,a} r_a {\Pi_0}^{z \bar z}_{a i}(T,t) {\Pi_0}^{z z}_{i j}(t,t) \left(r_i - r_j\right) \\
        &= -\frac{1}{N^2} \int_0^T dt \, \sum_{i,j,a} r_a {\Pi_0}^{z \bar z}_{a i}(T,t) \left({\Pi_0}^{z z}_{i j}(t,t) + N \delta_{ij} p_i(t)\right) r_j,
\end{align}
This result shows that the correction to the mean fitness  is
$\mathcal O(1/N)$
 the mean fitness in the limit of infinite population.
This result can be rewritten in a more revealing form. Let $\bar r(t)$ be a random variable defined as
\begin{align}
    \bar r(t) \equiv \frac{1}{N} \sum_i r_i \left(N_i(t) - \left<N_i(t)\right>\right)
\end{align}
in the limit of large population size. The finite population correction to the average fitness can then be written as
\begin{align}
    \left<\delta r(T)\right> &= - \int_0^T \left<\bar r(T) \, \bar r(t)\right> \, dt
\end{align}
and its time integral as
\begin{align}
    \int_0^T \left<\delta r(t)\right> \, dt = - \int_0^T dt \int_0^t dt' \, \left<\bar r(t) \bar r(t')\right>
        = - \frac12 \left<\left(\int_0^T \bar r(t) \, dt \right)^2\right>.
\end{align}
This expression for the average fitness correction, which resembles a fluctuation dissipation theorem, implies that the time-average of the finite-population shift is always negative. In other words, the average fitness of a large finite population is smaller than that of a population of infinite size. Note that this result is perturbative, valid for large population size $N$, and it does not require the average fitness to be a monotonic function of $N$ for small $N$. On complex fitness landscapes, it is possible for small asexual populations to achieve a higher average fitness than larger ones \cite{Jain2011}. Nonetheless, for sufficiently large population sizes, the time-integrated average fitness increases monotonously with population size.

\section{The Landscape}
\label{sec:landscape}

\begin{figure}[htbp]
    \begin{center}
        \includegraphics{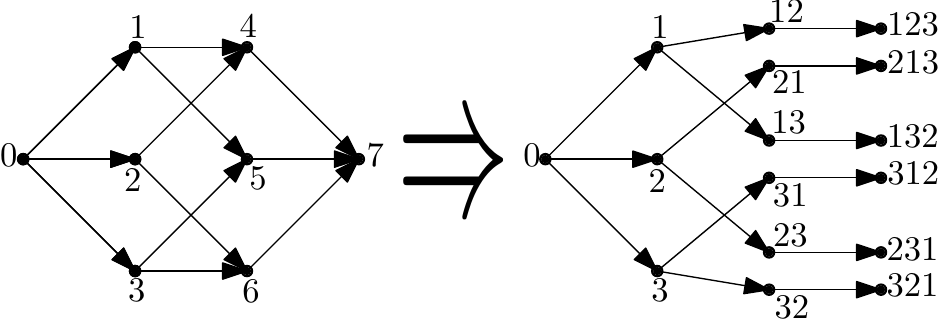}
    \end{center}
    \caption{Left-hand side: the state-space for a fitness landscape with three forward-mutations and no back-mutations.  Each node, $i$, is a particular genotype.  The replication rate of each genotype is $r_i$. Right-hand side (discussed in Section \ref{paths}): The state-space can be expanded to include mutational histories. Each two-mutation state is split into 2! = 2 states while the three-mutation state is split into 3! = 6 states.  The node is now identified by a vector which conveys the mutational history of a particular path through the landscape.}
    \label{fig:expandStateSpace}
\end{figure}

The analytical expressions developed in this paper are applicable to arbitrary fitness landscapes and mutational pathways. However, we now describe in some detail the implications for fitness landscapes \cite{Cowperthwaite2007} defined by a certain number of fitness loci $l$ with two alleles each. Genotypes that differ from each other by exactly one point mutation in one of the loci are connected in the mutation matrix. Each position in sequence space is thus connected by a mutation event to $l$ other genotypes. Figure \ref{fig:expandStateSpace} shows the geometry of the landscape for the case of three loci.  Typically in this landscape, the fitness of each state increases upon moving to the right in the figure.

\section{Fluctuations around the Mean}
\label{sec:fluctuations}
\begin{figure}[htbp]
    \begin{center}
        \subfloat[]{\includegraphics[width=0.32\textwidth]{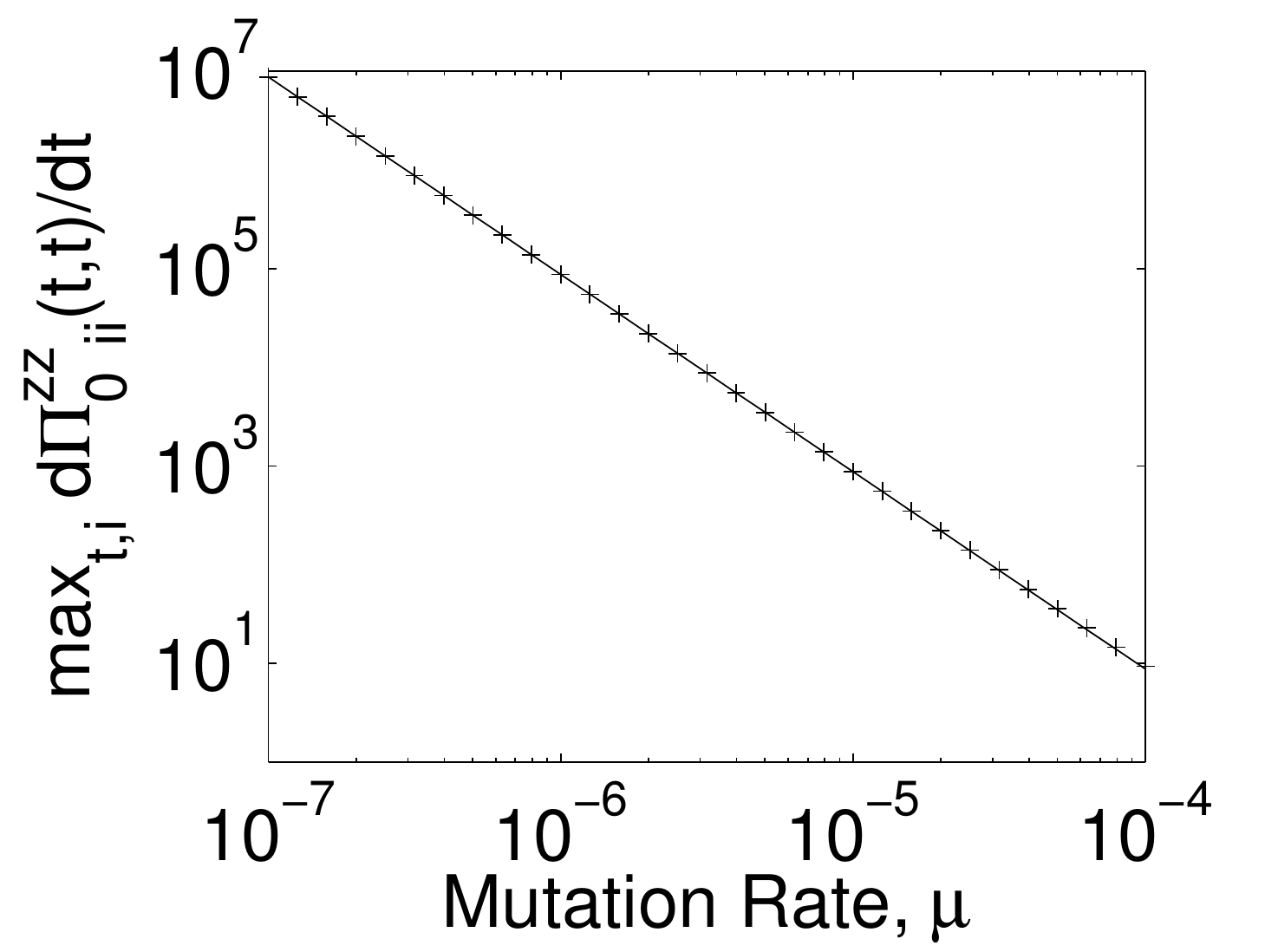}}
        \subfloat[]{\includegraphics[width=0.32\textwidth]{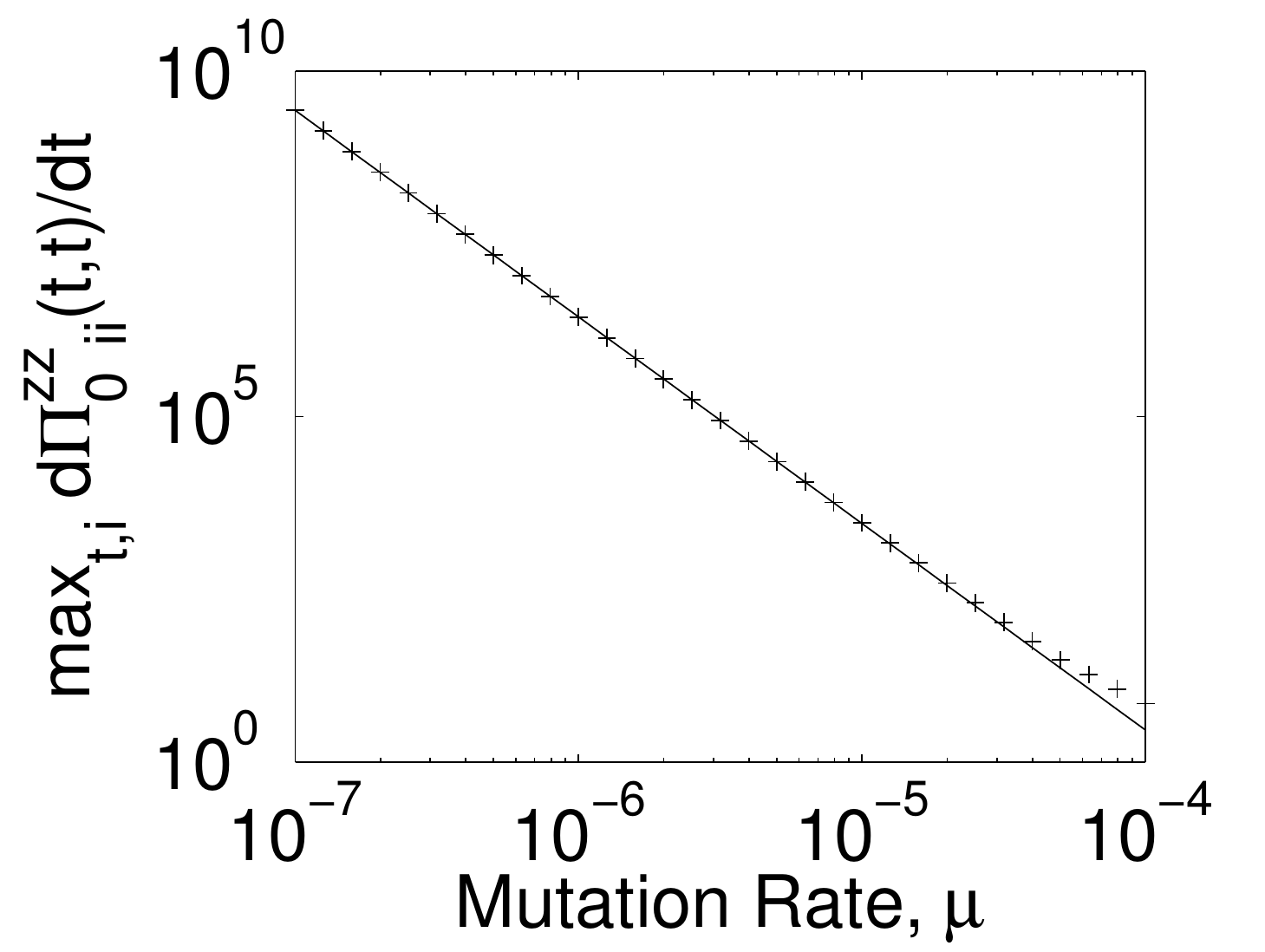}}
        \subfloat[]{\includegraphics[width=0.32\textwidth]{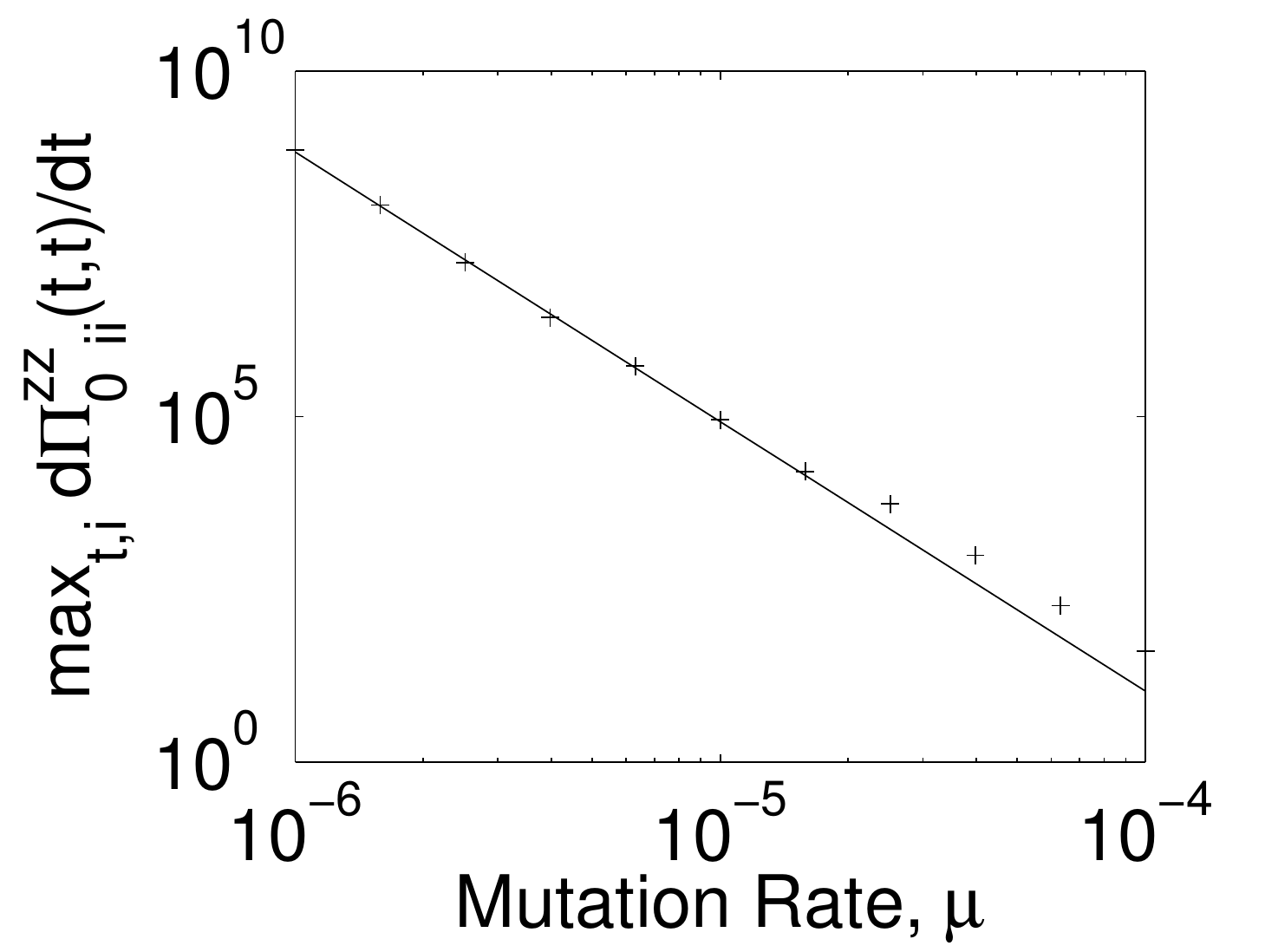}}
    \end{center}
    \caption{The maximal change of the variance with time (+), i.e. $\max_{t,i} d{\Pi_0}^{zz}_{ii}(t,t)/dt$ where ${\Pi_0}^{zz}$ is obtained from Eqs.\ \ref{eq:diffEqPizz} and \ref{eq:initCondPizz}, depends on the mutation rate as an inverse power law. Shown are calculations for a non-epistatic version of the landscape as described in section \ref{sec:landscape} with a) two possible mutations  --- $r_0 = 0, \Delta r_1 \approx 0.049, \Delta r_2 \approx 0.010$, b) three possible mutations --- $r_0 = 0, \Delta r_1 \approx 0.049, \Delta r_2 \approx 0.010, \Delta r_3 \approx 0.002$ --- and c) four possible mutations  --- $r_0 = 0, \Delta r_1 \approx 0.049, \Delta r_2 \approx 0.020, \Delta r_3 \approx 0.006, \Delta r_4 \approx 0.002$.  In this case, the fitness of each state is simply the sum of contributions from each mutation.
The solid lines indicate power law fits using the values for $\mu \le 10^{-5}$. Their exponents are a) -1.999, b) -2.989, and c) -3.939.
The exponent is observed to be equal to the number of mutational steps in the landscape.
}
    \label{fig:corrVsMu2Mut}
\end{figure}

The matrices ${\Pi_0}^{z z}(t,t)$ and ${\Pi_0}^{z \bar z}(T,t)$ can be understood intuitively. In the limit of large $N$, the off-diagonal elements of ${\Pi_0}^{z z}(t,t)$ describe the covariances between the occupation numbers at time $t$ while the diagonal elements are related to the variances of the occupation numbers at time $t$ by
\begin{align}
    \label{eq:variances}
    \frac{1}{N^2} (\delta N_a(t))^2 \sim \frac{1}{N} \left(p_a(t) + \frac{1}{N} {\Pi_0}^{z z}_{a a}(t,t)\right).
\end{align}
At different times, ${\Pi_0}^{z z}(T,t)$ gives the cross-covariances between the occupation numbers at times $T$ and $t$. The matrix ${\Pi_0}^{z \bar z}_{a i}(T,t)$ relates the correlations at different times to the same-time correlations via
\begin{align}
    {\Pi_0}^{z z}(T,t) = {\Pi_0}^{z \bar z}(T,t) {\Pi_0}^{z z}(t,t).
\end{align}
We observe numerically that for small mutation rates, the fluctuations are proportional to a negative power of the mutation rate. Specifically,
\begin{align}
    \frac{1}{N^2} (\delta N_a(t))^2 \sim 
\frac{1}{N} \left(\frac{r}{\mu}\right)^k,
\label{36}
\end{align}
where $k$ is the number of mutational steps as shown in Fig.\ \ref{fig:corrVsMu2Mut}. This dependence can also be shown analytically for sufficiently simple landscapes. See section \ref{sec:muToTheMinusK} in the appendix for one example.
Thus the expansion, which naively appears to be in $1/N$ is actually in
$1/(N \mu^k)$.  Thus, the expansion breaks down when $\mu < 1/N^{1/k}$.
The expansion is valid for large $N$ and $\mu \gg 1/N^{1/k}$.

\begin{figure}[htbp]
    \subfloat[]{\includegraphics[width=0.5\textwidth]{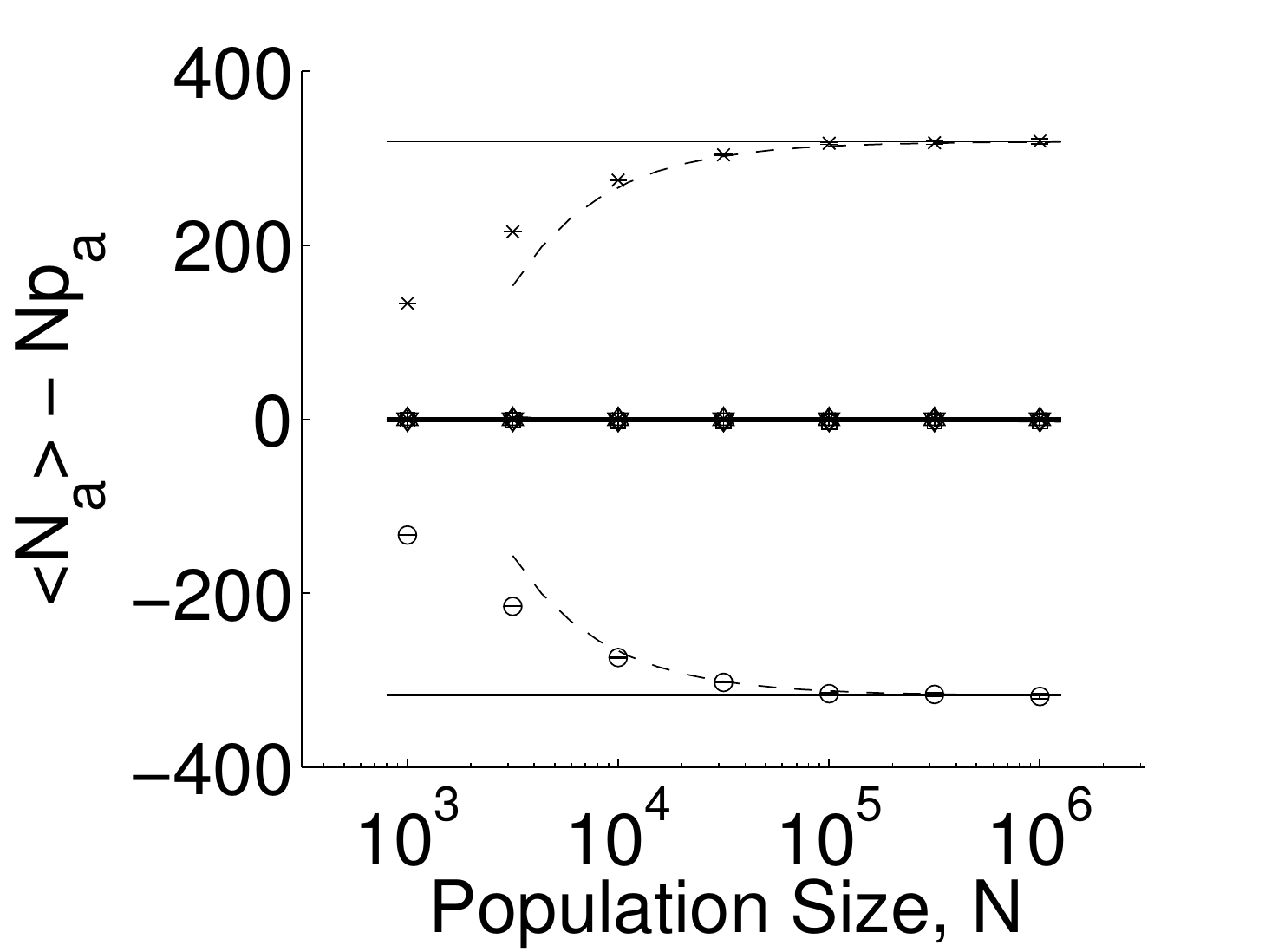}} \hfill
    \subfloat[]{\includegraphics[width=0.5\textwidth]{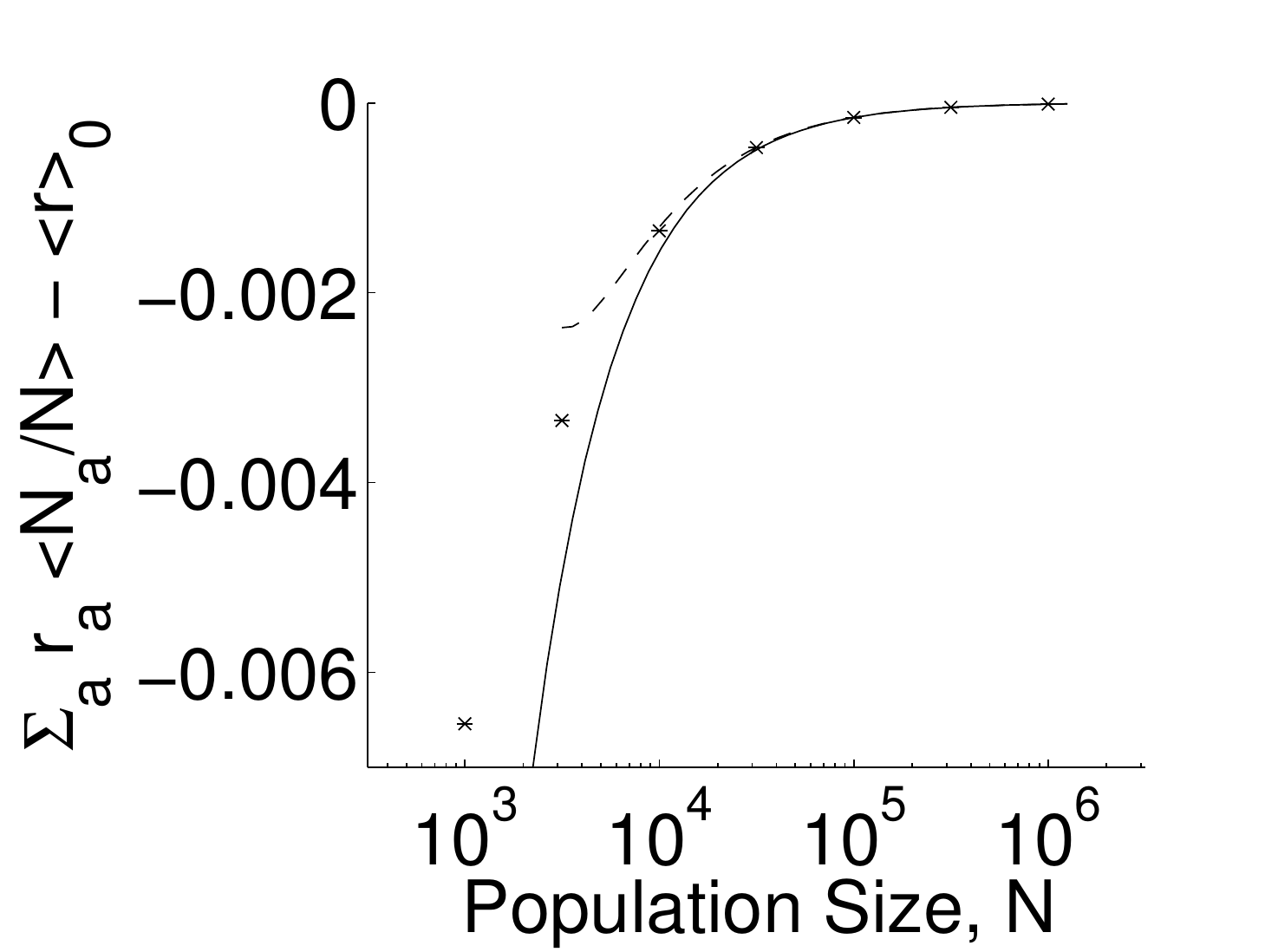}} 
    \caption{(a) Finite-population correction to the average occupation numbers (left-hand side of Eq.\ \ref{eq:firstCorrectionSim}) as a function of population size, $N$, on a three-mutation landscape as shown in Fig.\ \ref{fig:expandStateSpace} including back-mutations. Shown are data for a mutation rate of $\mu = 10^{-5}$ and replication rates of $r_0 = 0, r_1 \approx 0.049, r_2 \approx 0.010, r_3 \approx 0.002, r_4 \approx 0.059, r_5 \approx 0.051, r_6 \approx 0.012$, and $r_7 \approx 0.061$. The time is chosen as $T = 157.5$ which approximately maximizes $\left<N_0\right>(T) - N p_0(T)$. As $N$ increases, the corrections obtained from stochastic simulations --- $N_0(\times)$, $N_1(\bigcirc)$, $N_2(+)$, $N_3(\ast)$, $N_4(\Box)$, $N_5(\Diamond)$, $N_6(\bigtriangledown)$, $N_7(\bigtriangleup)$ --- converge to the values predicted by the theory (solid lines). The dashed curves show the second order expansion, given by Eqs.\ \ref{eq:firstCorrectionSim} and \ref{eq:secondCorrection}. The error bars are one standard error. 
(b) Finite-size correction to the mean population fitness.
The average replication rate in the population is linear in the occupation
numbers, being equal to $\frac1N \sum_i r_i N_i(t)$, and so it too converges to
the quasispecies result in the limit of a large population.}
    \label{fig:correctionVsN}
\end{figure}

\begin{figure}[htbp]
    \begin{center}
        \includegraphics[width=0.6\textwidth]{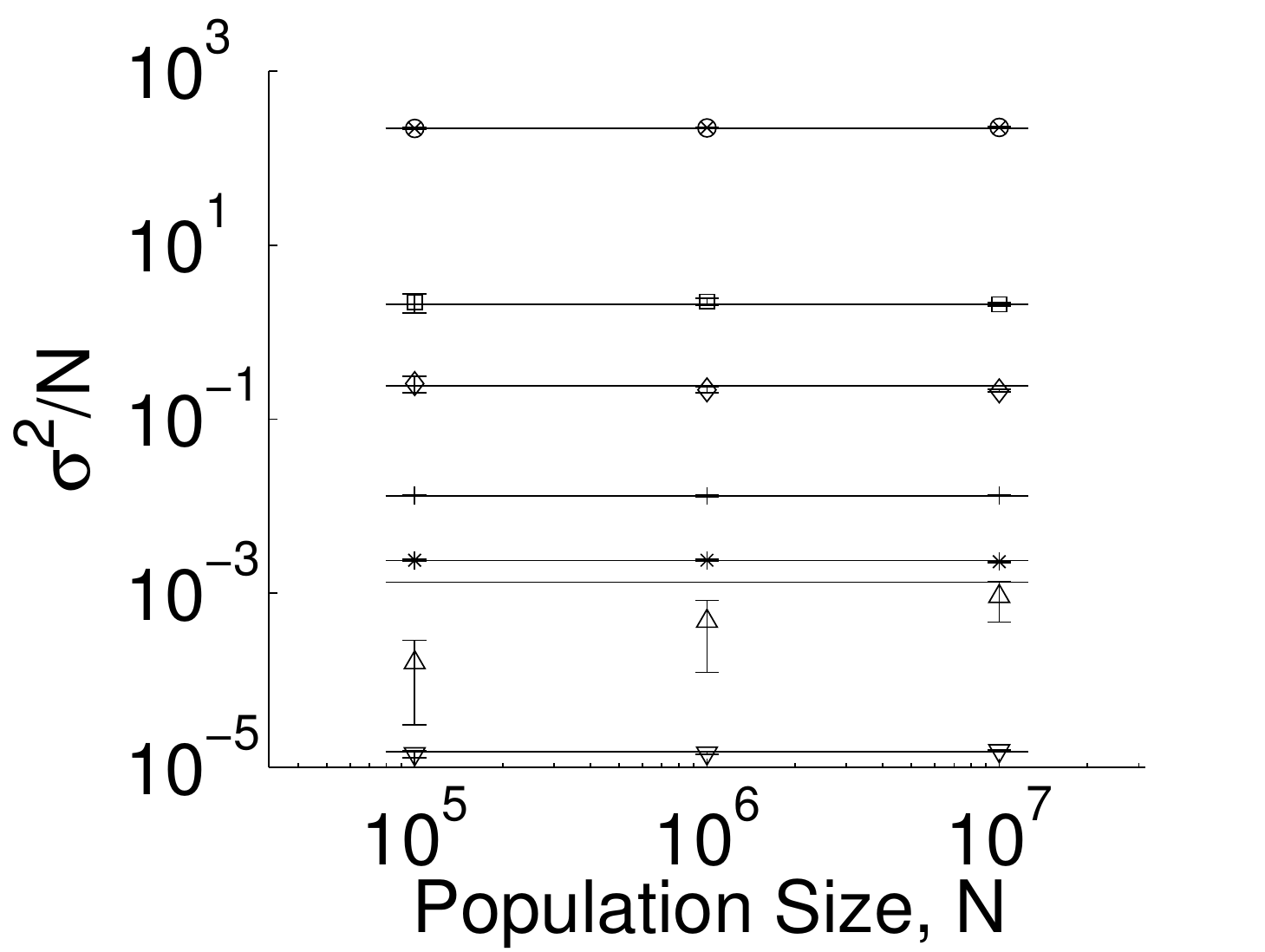}
    \end{center}
    \caption{Variances divided by population size, $N$ as a function of $N$. The values obtained from stochastic simulations --- $N_0(\times)$, $N_1(\bigcirc)$, $N_2(+)$, $N_3(\ast)$, $N_4(\Box)$, $N_5(\Diamond)$, $N_6(\bigtriangledown)$, $N_7(\bigtriangleup)$ --- agree with the values predicted by Eq.\ \ref{eq:variances} (solid lines). The time and other parameters are the same as in Fig.\ \ref{fig:correctionVsN}. The error bars are one standard error.}
    \label{fig:varianceVsN}
\end{figure}

\label{sec:compareToSimulations}
We verify our analytical results by performing stochastic simulations using the Lebowitz/Gillespie algorithm \cite{Bortz1975,Gillespie1976}. Rewriting Eq.\ \ref{eq:firstCorrection} for the first order shifts to the occupation numbers,
\begin{align}
    \label{eq:firstCorrectionSim}
    \left<N_a\right>(T) - N p_a(T) \sim \frac{1}{N} \int_0^T dt \, \sum_{i,j} {\Pi_0}^{z \bar z}_{a i}(T,t) {\Pi_0}^{z z}_{i j}(t,t) \left(r_i - r_j\right),
\end{align}
we observe that the finite population correction converges to a constant value for large $N$.
The average replication rate in the population is linear in the occupation
numbers. It is equal to $\frac1N \sum_i r_i N_i(t)$.  Therefore, the
average replication rate also
converges to the quasispecies result in the limit of a large population.
That is, the average replication rate is equal to that of the infinite
population plus a correction that is of order $1/N$ smaller.
Figure \ref{fig:correctionVsN} shows this convergence for one set of parameters. As a further check on our analytic results, we fit a cubic polynomial in $1/N$ to the simulation data displayed in Fig.\ \ref{fig:correctionVsN}. For the particular fitness parameters chosen here, the coefficients from this fit are $320.4 \pm 2.5$ for the constant term and $(-5.3 \pm 0.8) \times 10^5$ for the linear term, while our theory predicts $319.0$ and $-5.2 \times 10^5$, respectively. Here, the coefficient of the linear term is obtained from Eq.\ \ref{eq:secondCorrection} in Appendix \ref{sec:secondOrder}.
Similarly, we observe that the variances obtained from stochastic simulations agree with the analytic expression given in Eq.\ \ref{eq:variances} as shown in Fig.\ \ref{fig:varianceVsN}.


\section{Discussion and Conclusion}

\label{paths}
\begin{figure}[htbp]
    \begin{center}
        \includegraphics[width=0.6\textwidth]{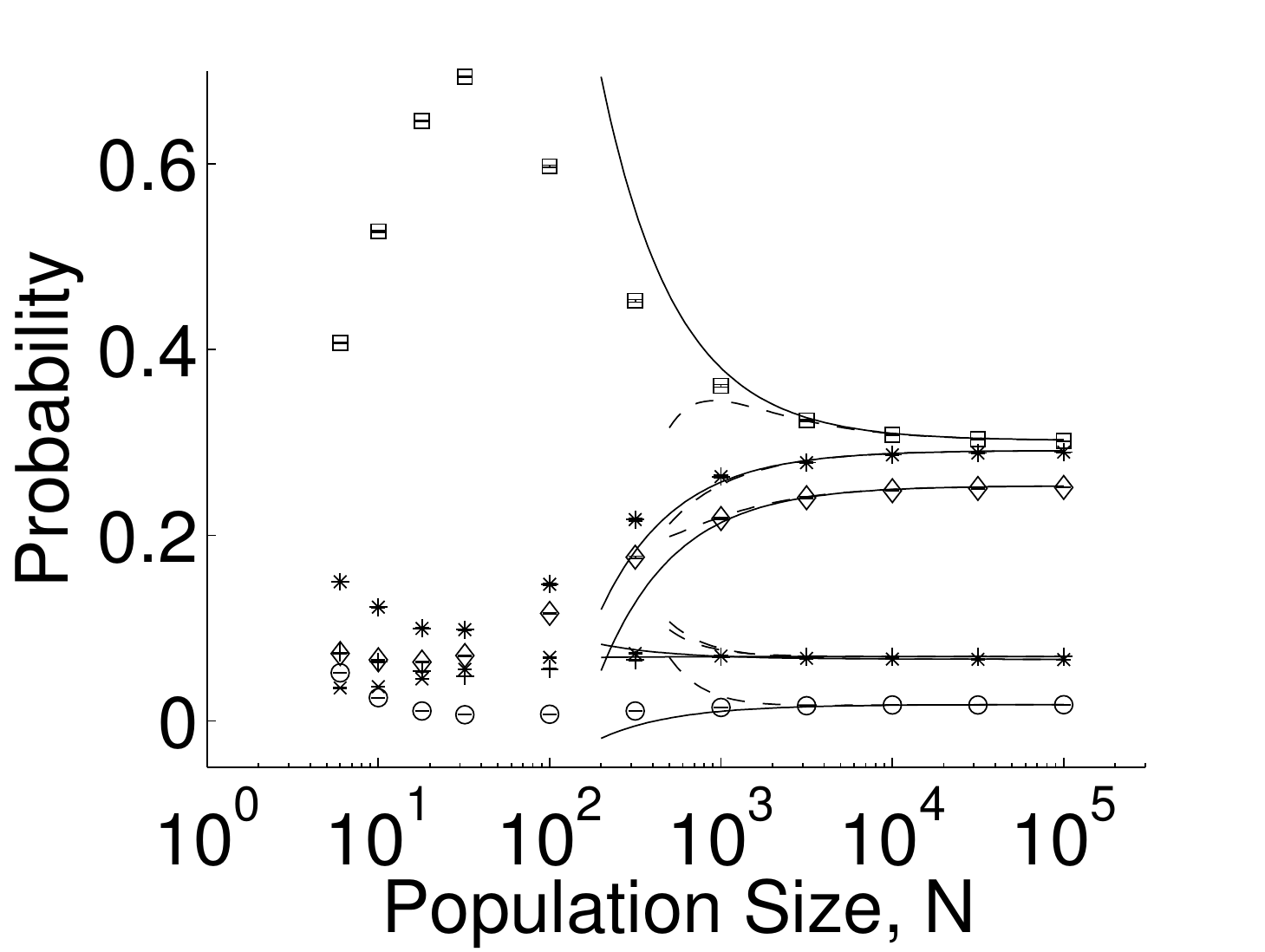}
    \end{center}
    \caption{Probability that a population will follow a certain mutational trajectory as a function of population size. Shown are data for the landscape in Fig.\ \ref{fig:expandStateSpace} excluding back-mutations with a mutation rate of $\mu = 10^{-3}$ and epistatic replication rates of $r_0 = 0, r_1 \approx 0.049, r_2 \approx 0.010, r_3 \approx 0.002, r_4 \approx 0.012, r_5 \approx 0.051, r_6 \approx 0.059$, and $r_7 \approx 0.061$. Equation \ref{eq:firstCorrectionSim} (solid lines) predicts the asymptotic behavior of the simulation values  --- $N_{123}(\times)$, $N_{213}(\bigcirc)$, $N_{132}(\Box)$, $N_{312}(+)$, $N_{231}(\ast)$, $N_{321}(\Diamond)$ ---  for large population sizes. The second order expansion (dashed lines) improves the prediction for sufficiently large populations. The error bars are one standard error.}
    \label{fig:pathPropabilitiesVsPopulationSize}
\end{figure}

Although the theory described in this paper was developed to study the time-evolution of the occupation numbers in sequence space, we can immediately apply these results to investigate which mutational paths individuals take. This allows us to predict the large $N$ behavior of the probability that a population will follow a certain mutational trajectory. To do this we simply expand the state space describing the identity of each individual to include not only the possible sequences but also the mutational histories. Figure \ref{fig:expandStateSpace} illustrates this expansion for the case of three mutations.   Figure \ref{fig:pathPropabilitiesVsPopulationSize} compares the probability of following a given path as obtained from stochastic simulations to the expressions given in Eqs.\ \ref{eq:firstCorrection} and \ref{eq:secondCorrection}. We again observe that the simulation results converge to the values predicted by the theory as the population size increases. Interestingly, we observe numerically that the probability for a population to take a certain mutational path varies with the population size in a non-monotonic fashion. In particular, there is an intermediate population size at which the population is most likely to take the dominant path through the landscape.

Fluctuations due to finite population can be quite large. As shown in Appendix \ref{sec:muToTheMinusK}, these fluctuations are proportional to an inverse power of the mutation rate.   That is, the expansion in $1/N$ has a coefficient that depends on a power of the inverse of the mutation rate.  For this reason, convergence to the infinite population limit can be exceedingly slow.  The coefficient in the expansion in $1/N$ also has a time dependence.  As shown in Appendix \ref{sec:largeTimeFluctuations}, this coefficient can be proportional to $t$, and so diverge at long times.
This divergence occurs when there are multiple final states, with equal
replication rates.
For example, the fluctuations diverge at long times in the expanded state space due to what may be termed fixation of path probabilities.


\label{sec:discussion}
In this paper we presented a path-integral formulation of evolution under a Moran-type process on arbitrary fitness landscapes. We derived analytic results that describe the dynamics exactly in the limit of an infinite population size and obtained an asymptotic expansion in the inverse of the population size for finite populations. We showed that the finite population correction to the time-averaged fitness is always negative, which implies that for sufficiently large population sizes the time-averaged fitness increases with population size. We also found that for small mutation rates, the infinite-population variances of the occupation numbers behave as $\mu^{-k}$ where $k$ is the number of mutational steps from the ancestral sequence. Finally, we showed how the formalism described in this paper can also be used to investigate which mutational path a population takes through the fitness landscape by expanding the sequence space to include mutational histories.

\section*{Acknowledgments}
This research was supported by the US National Institutes of Health
(1 R01 GM 100468--01).
JMP was also supported by the Basic Science Research Program
through the National Research Foundation of Korea funded
by the Ministry of Education, Science, and Technology
(grant number 2010--0009936).

\newpage
\appendix
\numberwithin{equation}{section}
\numberwithin{figure}{section}
\section{Second Order Correction}
\label{sec:secondOrder}
\begin{figure}[htbp]
    \subfloat[]{\includegraphics{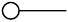}} \hfill
    \subfloat[]{\includegraphics{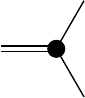}} \hfill
    \subfloat[]{\includegraphics{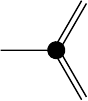}} \hfill
    \subfloat[]{\includegraphics{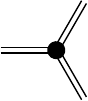}} \hfill
    \subfloat[]{\includegraphics{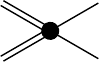}}
    \caption{Vertices for the diagrammatic expansion. A white circle represents an open time, while black circles stand for times that are integrated over.}
    \label{fig:vertices}
\end{figure}

\begin{figure}[htbp]
    \begin{center}
        \includegraphics{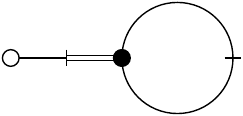}
    \end{center}
    \caption{Diagram for the $\mathcal O(N^0)$ correction to the average occupation numbers.}
    \label{fig:firstCorrDiag}
\end{figure}

\begin{figure}[htbp]
    \begin{center}
        \subfloat[Multiplicity: 2]{\includegraphics{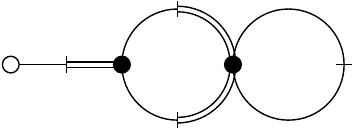}} \\
        \subfloat[Multiplicity: 1]{\includegraphics{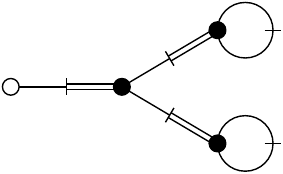}} \hspace{0.05\textwidth}
        \subfloat[Multiplicity: 2]{\includegraphics{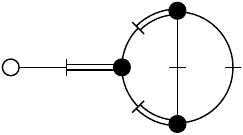}} \\
        \subfloat[Multiplicity: 4]{\includegraphics{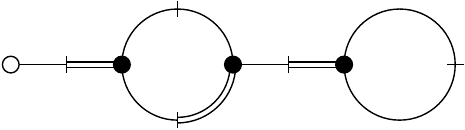}} \hspace{0.05\textwidth}
        \subfloat[Multiplicity: 2]{\includegraphics{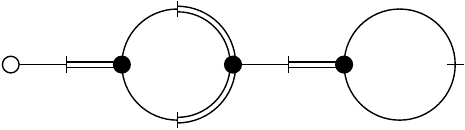}} \\
        \subfloat[Multiplicity: 8]{\includegraphics{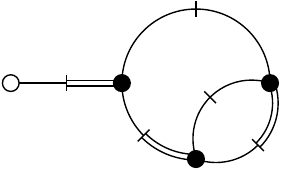}} \hspace{0.05\textwidth}
        \subfloat[Multiplicity: 4]{\includegraphics{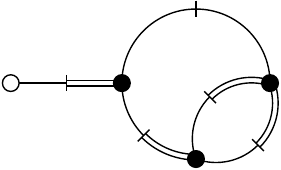}} \hspace{0.05\textwidth}
        \subfloat[Multiplicity: 8]{\includegraphics{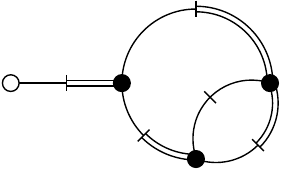}} \hspace{0.05\textwidth}
        \subfloat[Multiplicity: 12]{\includegraphics{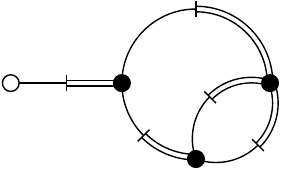}}
    \end{center}
    \caption{Diagrams for the $\mathcal O(N^{-1})$ correction to the average occupation numbers with their multiplicities.}
    \label{fig:secondCorrDiag}
\end{figure}

Equation \ref{eq:firstCorrection} gives the terms up $\mathcal O(N^{0})$ of an asymptotic expansion for the average occupation numbers in powers of $1/N$. We here determine the second order, $\mathcal O(N^{-1})$ correction terms. Figure \ref{fig:vertices} shows all possible vertices appearing in the diagrams.
Unlike the first correction term, which is derived from only the single non-vanishing diagram shown in Fig.\ \ref{fig:firstCorrDiag},
the second order correction term comes from the nine different diagrams shown in Fig.\ \ref{fig:secondCorrDiag}.
We obtain
\begin{align}
    \label{eq:secondCorrection}
    \frac{1}{N} \left<N_a\right>(T) \sim p_a(T)
        + \frac{1}{N^2} \int_0^T dt \, \sum_{i,j} {\Pi_0}^{z \bar z}_{a i}(T,t) {\Pi_0}^{z z}_{i j}(t,t) \left(r_i - r_j\right)
        + \frac{1}{N} \left<N_a\right>^{(2)}(T),
\end{align}
where
\begin{align}
    &\left<N_a\right>^{(2)}(T) \notag \\
    &=\frac{1}{2 N^2} \int_0^T dt \, \sum_{i,j} \left({\Pi_0}^{z \bar z}_{a i}(T,t) - {\Pi_0}^{z \bar z}_{a j}(T,t)\right)
            \left(r_i - r_j\right) \notag \\
        &\hspace{1.7in}
            \int_0^t dt' \, \sum_{i',j'} r_{i'}
            \left({\Pi_0}^{z \bar z}_{i i'}(t,t') - {\Pi_0}^{z \bar z}_{i j'}(t,t')\right) {\Pi_0}^{z \bar z}_{j i'}(t,t') {\Pi_0}^{z z}_{i' j'}(t',t') \notag \\
    &+\frac{1}{4 N^3} \int_0^T dt \, \sum_{i,j} \left({\Pi_0}^{z \bar z}_{a i}(T,t) - {\Pi_0}^{z \bar z}_{a j}(T,t)\right)
            \left(r_i - r_j\right) \notag \\
        &\hspace{1.2in}
            \int_0^t dt' \, \sum_{i',j'} \left({\Pi_0}^{z \bar z}_{i i'}(t,t') - {\Pi_0}^{z \bar z}_{i j'}(t,t')\right) \left(r_{i'} - r_{j'}\right)
                \int_0^{t'} dt'' \, \sum_{i'',j''} \left(r_{i''} - r_{j''}\right) \notag \\
        &\hspace{1.7in} \left[
                \left({\Pi_0}^{z \bar z}_{j i''}(t,t'') - {\Pi_0}^{z \bar z}_{j j''}(t,t'')\right)
                \left({\Pi_0}^{z z}_{i' j'}(t',t') {\Pi_0}^{z z}_{i'' j''}(t'',t'') \right. \right. \notag \\
        &\left.\hspace{3.6in} +
                    2 {\Pi_0}^{z z}_{i' j''}(t',t'') {\Pi_0}^{z z}_{j' i''}(t',t'')\right) \notag \\
        &\hspace{1.7in} +2
                \left({\Pi_0}^{z \bar z}_{i' i''}(t',t'') - {\Pi_0}^{z \bar z}_{i' j''}(t',t'')\right)
                \left({\Pi_0}^{z z}_{j j'}(t,t') {\Pi_0}^{z z}_{i'' j''}(t'',t'') \right. \notag \\
        &\left.\left.\hspace{3.6in} +
                    2 {\Pi_0}^{z z}_{j i''}(t,t'') {\Pi_0}^{z z}_{j' j''}(t',t'') \right) \right] \notag \\
    &+\frac{1}{N^3} \int_0^T dt \, \sum_{i,j}
                \left({\Pi_0}^{z \bar z}_{a i}(T,t) - {\Pi_0}^{z \bar z}_{a j}(T,t)\right) \left(r_i - r_j\right) \\
        &\hspace{0.68in}
                \int_0^t dt' \, \sum_{i',j'}
                \left({\Pi_0}^{z \bar z}_{i i'}(t,t') - {\Pi_0}^{z \bar z}_{i j'}(t,t')\right)
                \left(r_{i'} - r_{j'}\right)
                \int_0^{t'} dt'' \, \sum_{i'',j''} r_{i''} \notag \\
        &\hspace{1.0in}
                \left[{\Pi_0}^{z \bar z}_{j i''}(t,t'') \left({\Pi_0}^{z \bar z}_{j' i''}(t',t'') - {\Pi_0}^{z \bar z}_{j' j''}(t',t'')\right)
                    \left({\Pi_0}^{z z}_{i' i''}(t',t'') {z_c}_{j''}(t'') + {\Pi_0}^{z z}_{i' j''}(t',t'') {z_c}_{i''}(t'')\right) \right. \notag \\
        &\hspace{1.0in} +
                {\Pi_0}^{z \bar z}_{j' i''}(t',t'') \left({\Pi_0}^{z \bar z}_{j i''}(t,t'') - {\Pi_0}^{z \bar z}_{j j''}(t,t'')\right)
                    \left({\Pi_0}^{z z}_{i' i''}(t',t'') {z_c}_{j''}(t'') + {\Pi_0}^{z z}_{i' j''}(t',t'') {z_c}_{i''}(t'')\right) \notag \\
        &\hspace{1.0in} + \left.
                {\Pi_0}^{z \bar z}_{j' i''}(t',t'') \left({\Pi_0}^{z \bar z}_{i' i''}(t',t'') - {\Pi_0}^{z \bar z}_{i' j''}(t',t'')\right)
                    \left({\Pi_0}^{z z}_{j i''}(t,t'') {z_c}_{j''}(t'') + {\Pi_0}^{z z}_{j j''}(t,t'') {z_c}_{i''}(t'')\right) \right] \notag \\
    &+\frac{1}{N^3} \int_0^T dt \, \sum_{i,j} \left({\Pi_0}^{z \bar z}_{a i}(T,t) - {\Pi_0}^{z \bar z}_{a j}(T,t)\right)
                \left(r_i - r_j\right) \notag \\
        &\hspace{1.7in}
                \int_0^t dt' \, \sum_{i',j'} r_{i'}
                \int_0^{t'} dt'' \, \sum_{i'',j''} {\Pi_0}^{z z}_{i'' j''}(t'',t'') \left(r_{i''} - r_{j''}\right) \notag \\
        &\hspace{1.7in}
                {\Pi_0}^{z \bar z}_{i i'}(t,t') \left({\Pi_0}^{z \bar z}_{j i'}(t,t') - {\Pi_0}^{z \bar z}_{j j'}(t,t')\right)
                    \left({\Pi_0}^{z \bar z}_{i' i''}(t',t'') {z_c}_{j'}(t') + {\Pi_0}^{z \bar z}_{j' i''}(t',t'') {z_c}_{i'}(t')\right) \notag \\
    &+\frac{2}{N^2} \int_0^T dt \, \sum_{i,j} \left({\Pi_0}^{z \bar z}_{a i}(T,t) - {\Pi_0}^{z \bar z}_{a j}(T,t)\right)
                \left(r_i - r_j\right) \int_0^t dt' \, \sum_{i',j'} {\Pi_0}^{z \bar z}_{i i'}(t,t') \left(r_{i'} - r_{j'}\right) \notag \\
        &\hspace{1.7in}
                \sum_{i''}
                {\Pi_0}^{z \bar z}_{j i''}(t,0) {\Pi_0}^{z \bar z}_{i' i''}(t',0) {\Pi_0}^{z \bar z}_{j' i''}(t',0) n_{i''} (0) \notag
\end{align}

\section{Fluctuations Proportional to a Negative Power of the Mutation Rate}
\label{sec:muToTheMinusK}
In this appendix we consider a special case of the model described in section \ref{sec:mapToFieldTheory}, for which we show analytically that for small mutation rates, $\mu$, the variance in the infinite population occupation numbers is proportional to $N\mu^{-k}$, where $k$ is the number of mutational steps in the landscape. 
We work in the limit that $N \to \infty$.  We seek to understand when
the $1/N$ expansions of Eqs.\ (\ref{eq:firstCorrection})
and (\ref{eq:variances}) break down.
We will show that for small $\mu$,
 the naive expansion in $1/N$ is actually an expansion
in $1/(N \mu^k)$.  The expansions in  Eqs.\ (\ref{eq:firstCorrection})
and (\ref{eq:variances}), therefore, break down when 
 $\mu < 1/N^{1/k}$.
In other words, the expansion is valid for large $N$ and $\mu \gg 1/N^{1/k}$.
Let there be $k+1$ positions in sequence space linked by $k$ mutations which occur at equal rate $\mu$ such that $\Delta_{ij} = \delta_{i,j-1}$ for $i < k$, where $\delta_{i,j}$ is the Kronecker delta. The fitness increases in the direction of mutations (all mutations are beneficial) but the fitness increments decrease monotonically. This landscape is commonly encountered when there is a dominant path through a landscape. For example, we encountered this case when applying our theory to long-term experimental studies of bacterial evolution \cite{Azevedo}.
\begin{figure}[htbp]
    \begin{center}
        \includegraphics{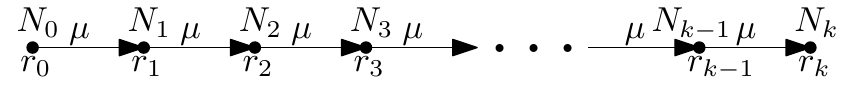}
    \end{center}
    \caption{A simple landscape in which mutations occur at rate $\mu$, without back mutation, the replication rate at position $i$ is $r_i$, and $N_i$ is the occupation number at position $i$.}
    \label{fig:setup}
\end{figure}
Fig.\ \ref{fig:setup} shows a graphical representation of this landscape. We assume that the mutation rate is very small, $\mu \ll r$, and that there is no back mutation. Initially, the entire population is in the starting state, $N_i(t=0) = N\delta_{i,0}$. For this simple landscape Eq.\ \ref{eq:diffEqP} can be solved explicitly for the infinite population occupation numbers. In the limit as $\mu \rightarrow 0$, we have
\begin{align}
 p &= \sum_{i=0}^{k} \sum_{b=0}^i \mu^i \gamma^i_b e^{r_b t} \label{eq:simplep0}\\
    p_i &= p^{-1} \sum_{b=0}^i \mu^i \gamma^i_b e^{r_b t}, \label{eq:simplepi}
\end{align}
where
\begin{align}
    \gamma^i_b = \begin{cases} \frac{\displaystyle (-1)^{i-b}}{\displaystyle \prod_{j=b+1}^i \left(r_j-r_b\right) \prod_{j=0}^{b-1} \left(r_b-r_j\right)}
            &\qquad b \le i \\
        0 &\qquad b>i. \end{cases}
\end{align}
Substitution into Eq.\ (\ref{eq:diffEqP}) confirms these solutions in the
$\mu \rightarrow 0$ limit.

\begin{figure}[htbp]
\begin{center}
\subfloat[]{\includegraphics[width=0.33\textwidth]{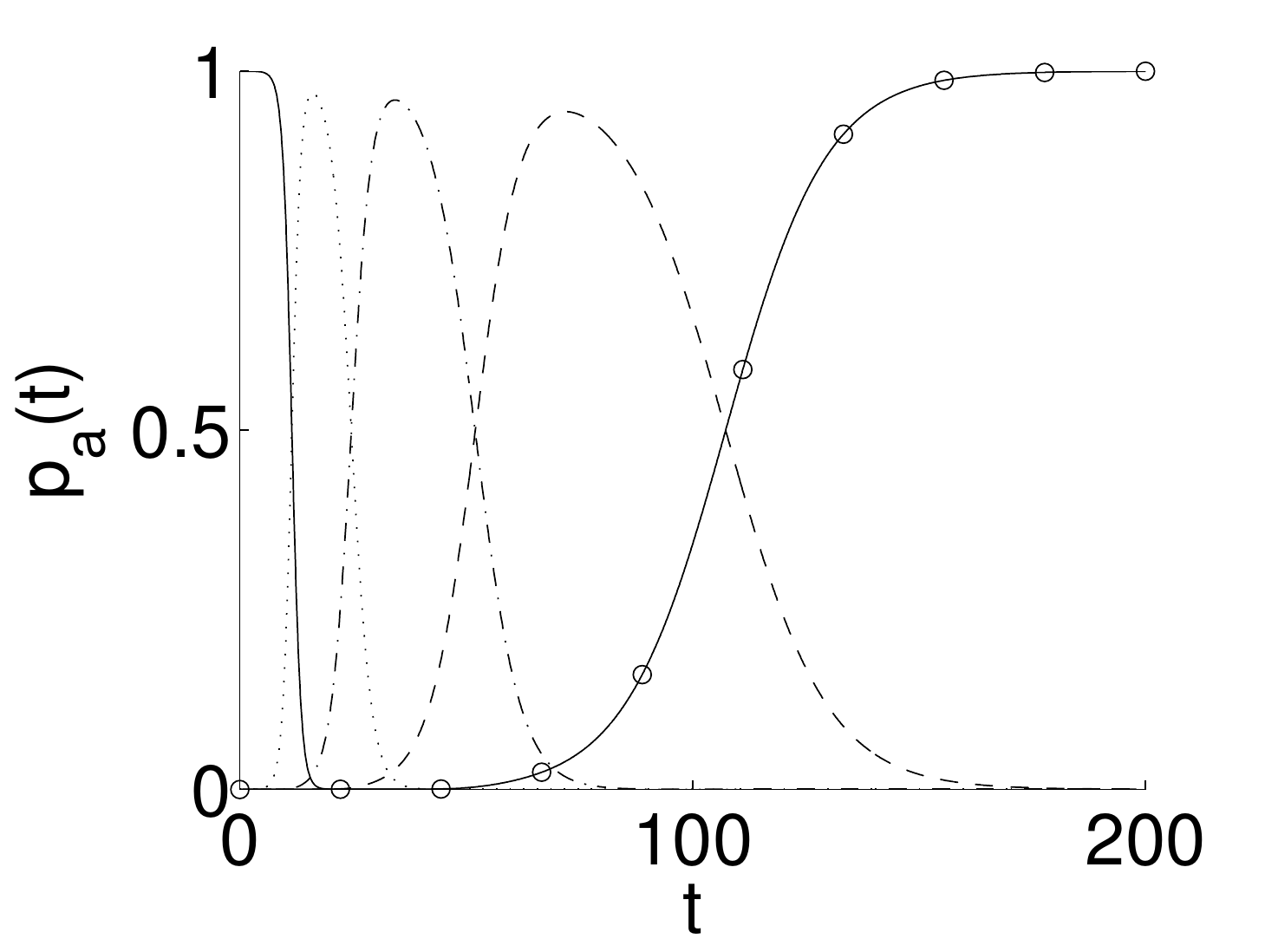}}
\hfill
\subfloat[]{\includegraphics[width=0.33\textwidth]{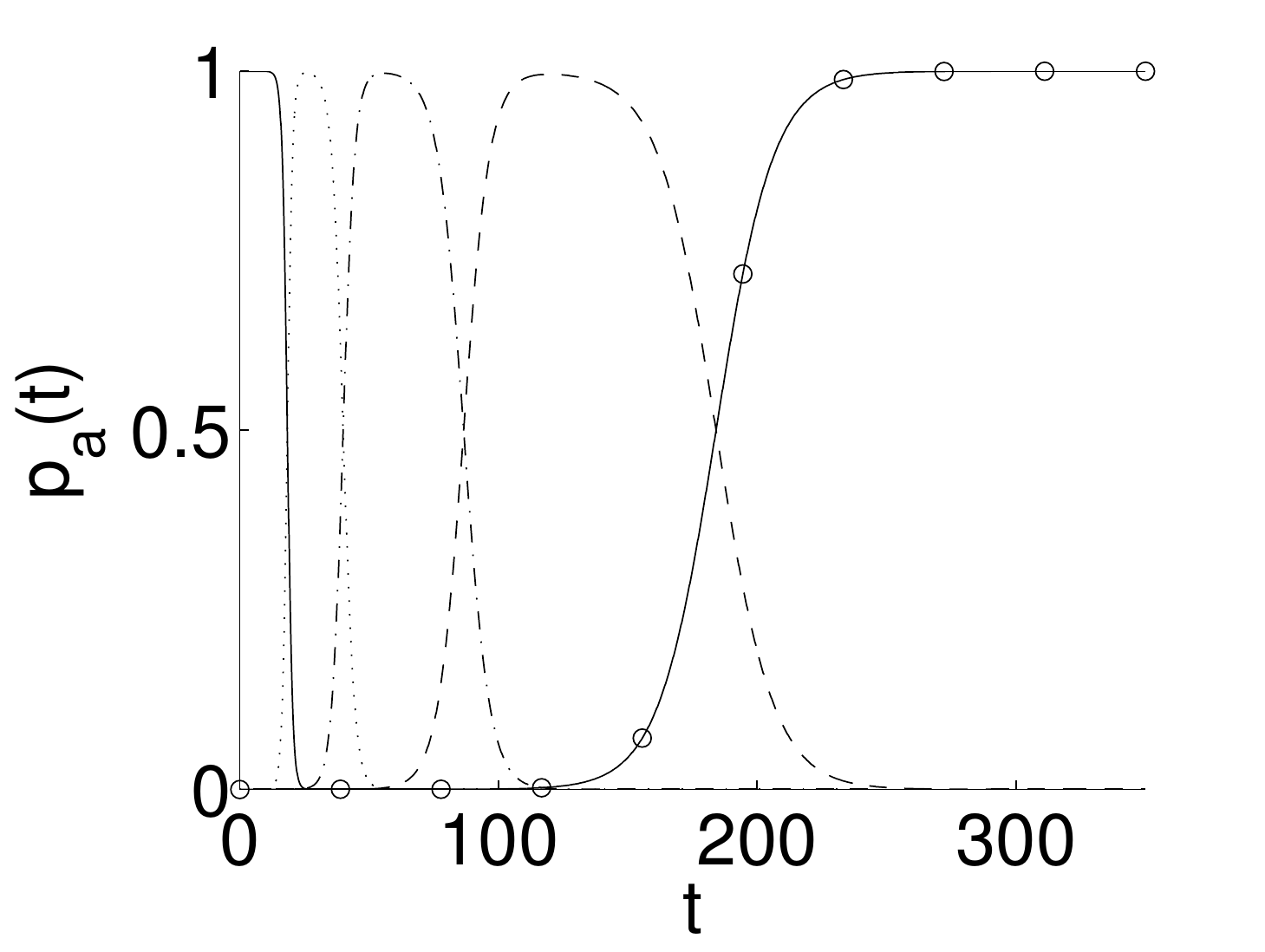}}
\hfill
\subfloat[]{\includegraphics[width=0.33\textwidth]{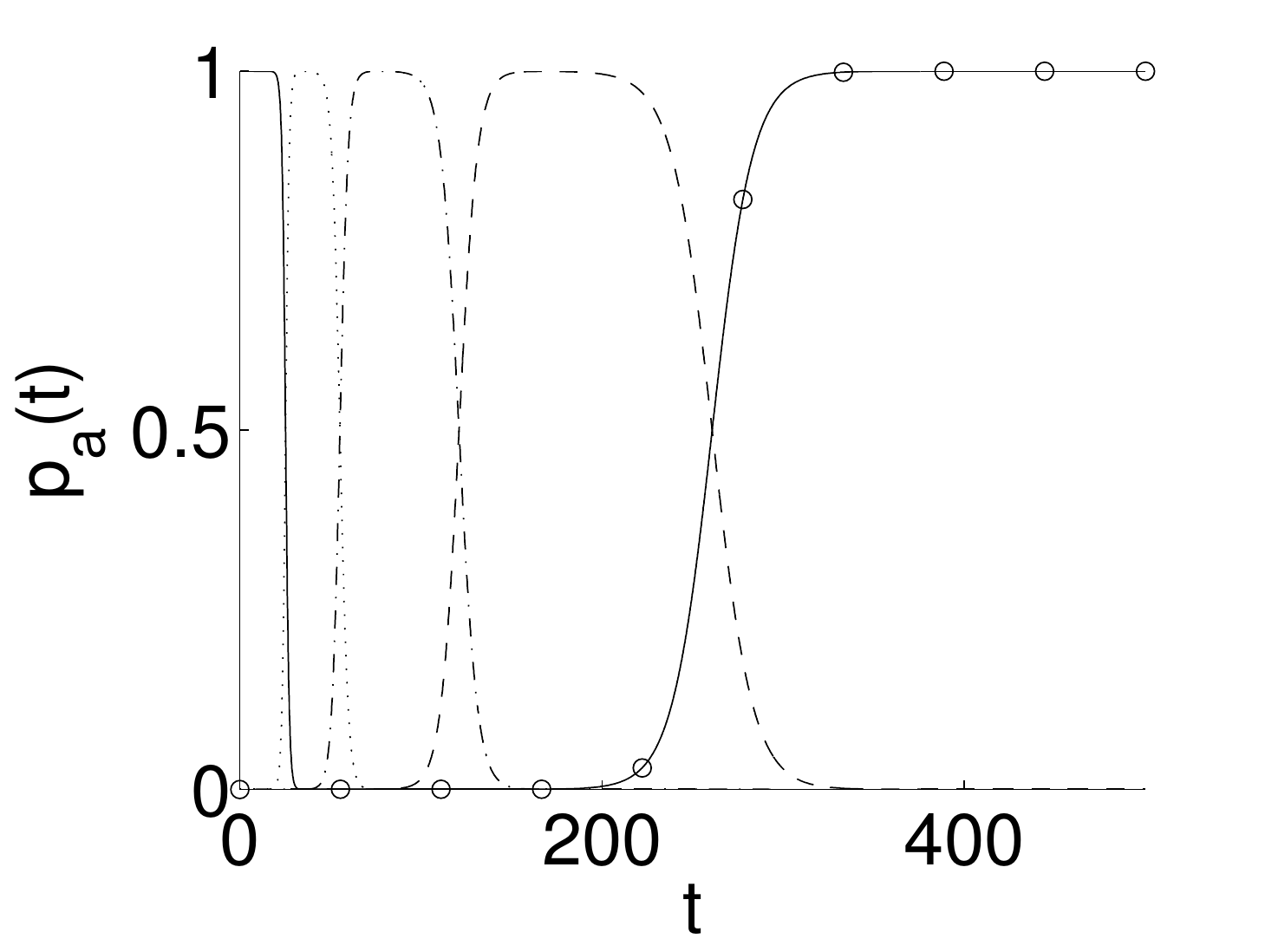}}
 \end{center}
 \caption{Infinite population occupation numbers versus time for $k =
4$, $r_0=0$, $r_1 = 1.00$, $r_2 = 1.45$, $r_3 = 1.65$, $r_4 = 1.74$, and three
different values for $\mu$: (a) $10^{-5}$, (b) $10^{-8}$, and (c)
$10^{-11}$. The occupation numbers, $p_0$ (solid), $p_1$ (dotted),
$p_2$ (dash-dotted), $p_3$ (dashed), $p_4$ (solid with circles), are
calculated using Eq.\ \ref{eq:closedFormP}. Note that as $\mu$ becomes
smaller, $p_a$ becomes more and more dominant during the time interval
$t_a < t < t_{a+1}$.}
 \label{fig:appendixMu5811p}
\end{figure}

\begin{figure}[htbp]
    \begin{center}
        \subfloat[]{\includegraphics[width=0.33\textwidth]{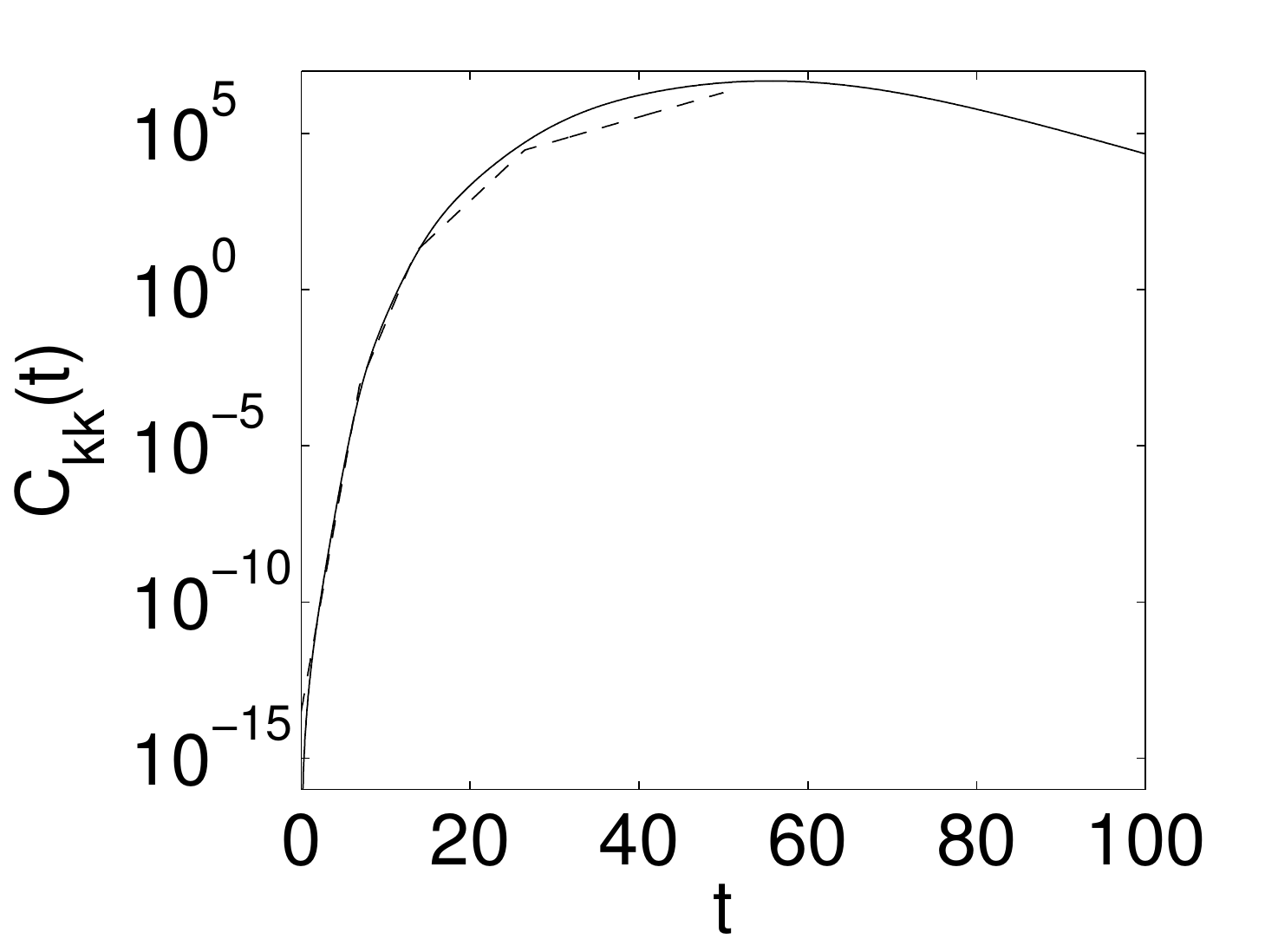}}
        \hfill
        \subfloat[]{\includegraphics[width=0.33\textwidth]{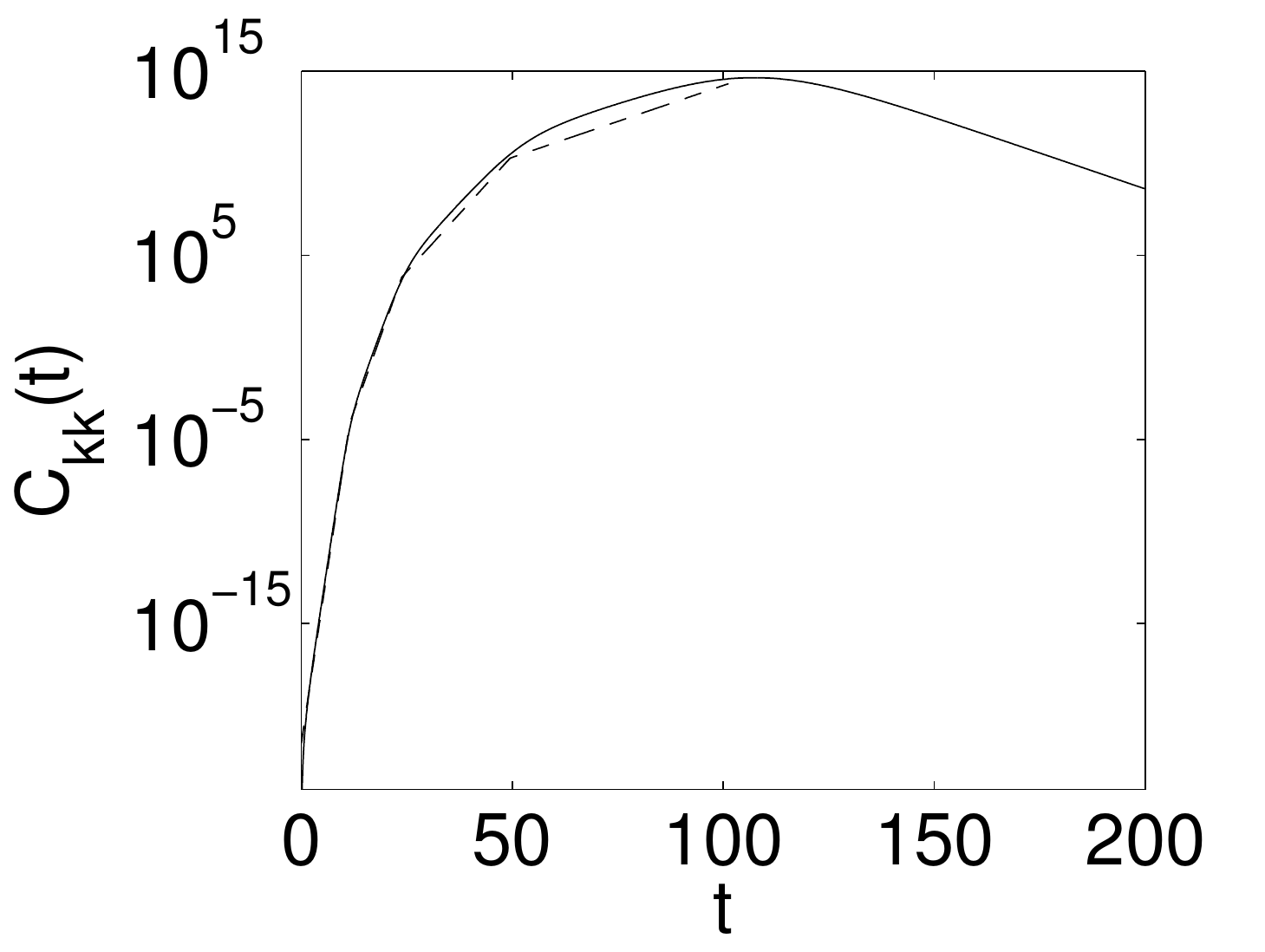}}
        \hfill
        \subfloat[]{\includegraphics[width=0.33\textwidth]{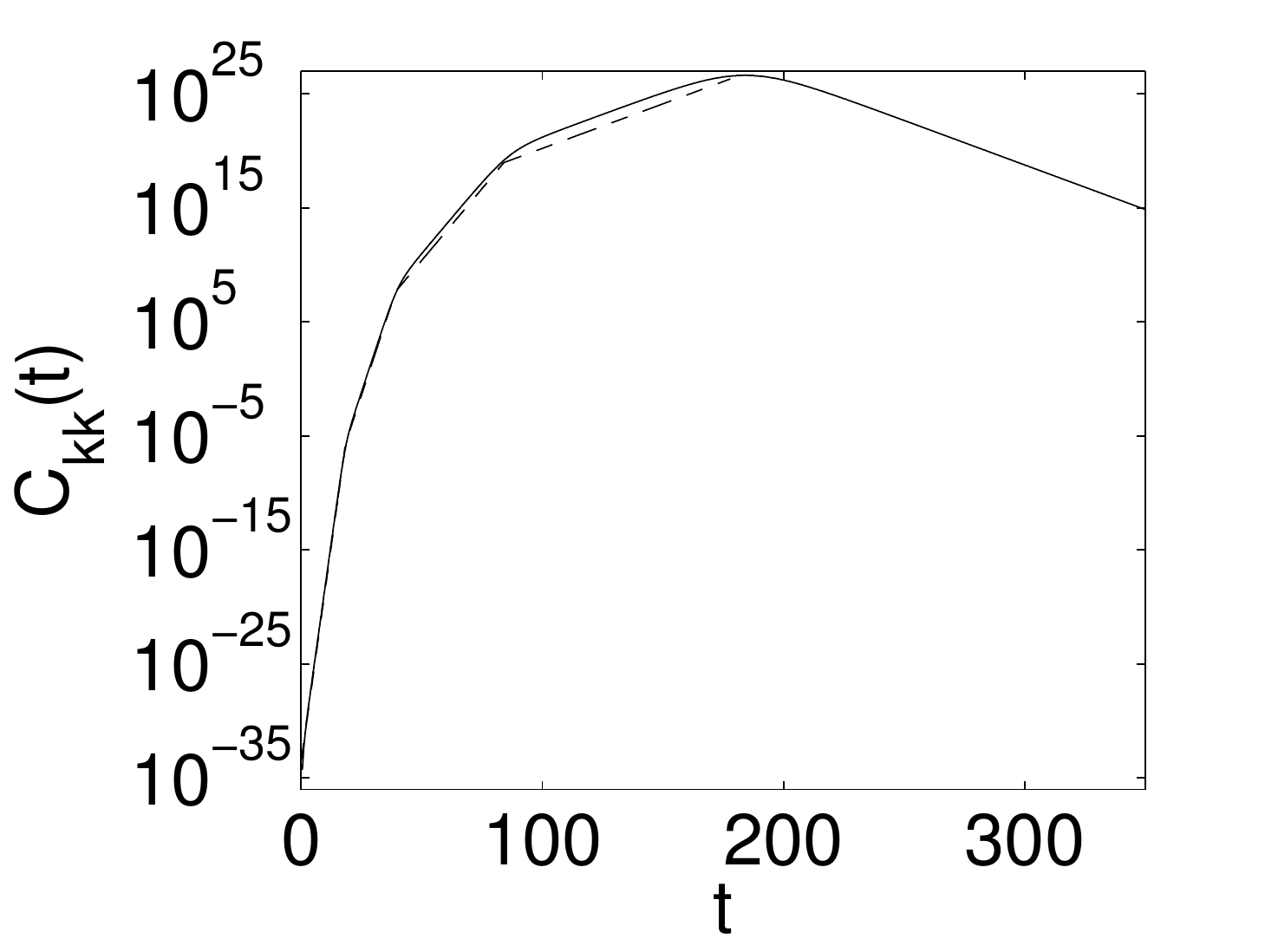}}
    \end{center}
    \caption{Infinite population variance of the final state vs. time for $k = 4$, $r_0 = 0$, $r_1 = 1.00$, $r_2 = 1.45$, $r_3 = 1.65$, $r_4 = 1.74$, and three different values for $\mu$: (a) $10^{-3}$, (b) $10^{-5}$, and (c) $10^{-8}$. Exact values calculated using Eq.\ \ref{eq:variances} (solid lines) and the approximation given in Eq.\ \ref{eq:CkkApprox} (dashed lines) are both shown. Note that $C_{kk}(t_k) \propto \mu^{-k}$.}
    \label{fig:appendixMu358}
\end{figure}

Let $C_{ij} \equiv \lim_{N \rightarrow \infty} \left(\left<N_i N_j\right> - \left<N_i\right>\left<N_i\right>\right) / N$ denote the infinite-population covariance matrix. From section \ref{sec:fluctuations} we know that
\begin{align}
    C_{ij}(t) = \delta_{ij} p_i(t) + \frac{1}{N} {\Pi_0}^{z z}_{i j}(t,t)
\end{align}
In the limit of infinite $N$, the correlation matrix $C$ converges to
a number independent of $N$.
We can show that
\begin{align}
    \label{eq:diffEqC}
    \frac{dC(t)}{dt} = \bar B(t) + A(t) C(t) + C(t) A^T(t),
\end{align}
with
\begin{align}
    \label{eq:initCondC}
    C_{ij}(0) = 0
\end{align}
and
\begin{align}
    \label{eq:Bbar}
    \bar B_{ij}(t) &=- \left(\mu \Delta_{ij} p_i(t) + \mu \Delta_{ji} p_j(t) + (r_i + r_j) p_i(t)p_j(t)\right) \notag \\
              & + \delta_{ij} \left(\mu \sum_a \Delta_{ai} p_a(t) + \mu \sum_a \Delta_{ia} p_i(t) + r_i p_i(t) + \expval{r} p_i(t)\right).
\end{align}
To compute $B$, one is allowed to use the infinite $N$ values for
$p_i(t)$ because finite $N$ corrections to $p_i(t)$ lead to higher order 
terms in the expansion Eq.\ (\ref{eq:variances}).
Let $t_0 = 0$, $t_a \equiv \ln\left(\Delta r_a/\mu\right)/\Delta r_a, 
0<a\le k$. 
We examine Eq.\ (\ref{eq:simplepi}). We consider $t > t_a$.
Expression \ref{eq:simplepi} for $p_a$ will be dominated by the last
term in the series, since the ratio of the magnitude of the
last term to the second to
last term is $ \exp(\Delta r_a t ) 
\prod_{j=0}^{j=a-2} (r_{a-1} - r_j) /
(r_{a} - r_j)=
(\Delta r_a / \mu) \exp[\Delta r_a (t - t_a)] 
\prod_{j=0}^{j=a-2} (r_{a-1} - r_j) / (r_{a} - r_j)$,
 and this is large for small $\mu$
and $t > t_a$.
Furthermore, the ratio of $p_a$ to $p_{a-1}$ is 
$(\mu / \Delta r_a) \exp(\Delta r_a t)
\prod_{j=0}^{j=a-2} (r_{a-1} - r_j) / (r_{a} - r_j)
 =
\exp[\Delta r_a (t - t_a)]
\prod_{j=0}^{j=a-2} (r_{a-1} - r_j) / (r_{a} - r_j) 
$, which is also large
for $t>t_a$.
The time interval from $t_a$ to
$t_{a+1}$ gets larger as $\mu$ gets smaller, so that the time
period during which
$p_{a-1}$ and $p_a$ are of similar magnitude, $t \sim t_a$, becomes less and
less significant.
Figure \ref{fig:appendixMu5811p} shows this result numerically.
Finally, the ratio of 
$p_{a+1}$ to $p_a$ is 
$\exp[\Delta r_{a+1} (t - t_{a+1})]
\prod_{j=0}^{j=a-1} (r_{a} - r_j) / (r_{a+1} - r_j) 
$, which is 
small for $t<t_{a+1}$.
Thus, for small $\mu$, in the time interval $t_a$ to $t_{a+1}$, 
most of the population is in state $a$.   That is, 
\begin{align}
    p_a(t) \gg p_{a'}(t) \qquad a' \ne a, t_a < t < t_{a+1}, \mu \rightarrow 0.
\end{align}
Using this result and keeping the lowest order in $\mu$ in Eq.\ \ref{eq:matA},
we find 
\begin{align}
    A_{ij}(t) &\sim \left(r_j - r_a\right)\left(\delta_{i,j} -  \delta_{i,a}\right), \qquad\qquad t_a < t < t_{a+1}, \mu \rightarrow 0
\end{align}
such that
\begin{align}
    \frac{dC_{ij}(t)}{dt} \sim \bar B_{ij}(t) + \left(r_i + r_j - 2 r_a\right) C_{ij}(t)
        - \sum_n \left(r_n - r_a\right) \left(\delta_{j,a} C_{i,n} + \delta_{i,a} C_{j,n}\right). \label{eq:simpleDiffEqC}
\end{align}
For this landscape, Eq.\ \ref{eq:Bbar} reduces to
\begin{align}
    \bar B_{ij}(t) &=- \left[\mu \delta_{i,j-1} p_i(t) + \mu \delta_{i,j+1} p_j(t) + (r_i + r_j) p_i(t) p_j(t)\right] \notag \\
              & + \delta_{ij} \left(\mu p_{i-1}(t) + \mu \left(1-\delta_{i,k}\right) p_i(t) + r_i p_i(t) + \expval{r} p_i(t)\right)
\end{align}
and, in particular,
\begin{align}
    \bar B_{kk}(t) &=- 2r_k \left(p_k(t)\right)^2 + \mu p_{k-1}(t) + r_k p_k(t) + \expval{r} p_k(t).
\end{align}
Substituting Eqs.\ \ref{eq:simplep0} and \ref{eq:simplepi} into this expression and keeping only the lowest power of $\mu$, we obtain, for $t < t_1$,
\begin{align}
    \bar B_{kk}(t < t_1) \sim r_k \mu^k \sum_{a=0}^k \gamma^k_a e^{r_a t} \qquad\qquad \mu \rightarrow 0
\end{align}
and thus
\begin{align}
    \frac{dC_{kk}(t)}{dt} \sim r_k \mu^k \sum_{a=0}^k \gamma^k_a e^{r_a t} + 2 r_k C_{kk}(t), \qquad\qquad t < t_1, \mu \rightarrow 0.
\end{align}
Integrating and only keeping terms to lowest order in $\mu$ yields
\begin{align}
    C_{kk}(t_1) \sim \mu^k \left(\frac{\Delta r_1}{\mu}\right)^{2r_k/\Delta r_1} \sum_{a = 0}^k \frac{\gamma^k_a}{2 - r_a/r_k}  \qquad\qquad \mu \rightarrow 0.
\end{align}
For later time periods, 
the evolution of $C_{kk}(t_1<t<t_k)$ is dominated by the second term in Eq.\ \ref{eq:simpleDiffEqC}
as $\mu \rightarrow 0$:
\begin{align}
    \frac{dC_{kk}(t)}{dt} \sim 2 \left(r_k - r_a\right) C_{kk}(t), \qquad\qquad t_a < t < t_{a+1}, 0 < a < k, \mu \rightarrow 0
\end{align}
with solution
\begin{align}
    \label{eq:CkkApprox}
    C_{kk}(t) \sim \mu^k e^{2 (r_k-r_a) t} \prod_{j = 1}^a \left(\frac{\Delta r_j}{\mu}\right)^{2} \sum_{a' = 0}^k \frac{\gamma^k_{a'}}{2 - r_{a'}/r_k} \notag \\
        t_a < t < t_{a+1}, 0 < a < k, \mu \rightarrow 0.
\end{align}
Fig.\ \ref{fig:appendixMu358} shows the convergence of this approximation to Eq.\ \ref{eq:variances} as $\mu \rightarrow 0$ for one set of replication rates.

Using Eq.\ \ref{eq:CkkApprox}, we find that as $\mu \rightarrow 0$
\begin{align}
    C_{kk}(t_k)
        &\sim \mu^k e^{2 \Delta r_k t_k} \sum_{a = 0}^k \frac{\gamma^k_{a}}{2 - r_{a}/r_k} \prod_{j = 1}^{k-1} \left(\frac{\Delta r_j}{\mu}\right)^{2} \notag \\
        &= \mu^{-k} \sum_{a = 0}^k \frac{\gamma^k_{a}}{2 - r_{a}/r_k} \prod_{j = 1}^{k} \left(\Delta r_j\right)^{2}.
\end{align}
The maximum of $C_{kk}(t)$ occurs near $t_k$.
This result follows from Eq.\ (\ref{eq:simpleDiffEqC}). The first term on
the righthand side of Eq.\ (\ref{eq:simpleDiffEqC}) only
matters during $0 < t < t_1$. After
that, $B_{kk}$ has a larger power of $\mu$ then $C_{kk}$ does.
The second term on the righthand side
is zero for $t > t_k$.  Thus, for $t> t_k$, only the
third term on the righthand side matters,  and it is negative.
Thus, for $t > t_k$, $C_{kk}(t)$ decreases.  It is for this reason
that the dashed curves in Fig. \ref{fig:appendixMu358} 
are shown for $0 < t < t_k$ only.

\section{Fluctuations in the Expanded State Space at Large Times}
\label{sec:largeTimeFluctuations}
Consider the expanded state space of a landscape as shown in Fig.\ \ref{fig:expandStateSpace} generalized to an arbitrary number of loci. For any finite population size $N$, the only sinks are the final states in which all mutations have occurred in some order, all of which have the same replication rate. Thus, after a certain time $t_f$, the occupation numbers at positions prior to the final states can be neglected so that the dynamics can be described by Eq.\ \ref{eq:masterEquation} with a single replication rate $r$ and without mutation,
\begin{align}
    \notag
    \frac{\partial}{\partial t} P(\boldsymbol N; t)
        &= \frac{r}{N} \sum_{i, j \ne i} \left[\left(N_i - 1\right) \left(N_j + 1\right) P(\boldsymbol N - \boldsymbol e_i + \boldsymbol e_j; t)
                - N_i N_j P(\boldsymbol N; t)\right].
\end{align}
From this we obtain that the average occupation numbers remain constant
\begin{align}
    \left<N_a(t)\right> = \text{const} = \left<N_a(t_f)\right> \qquad t \ge t_f
\end{align}
and that the covariances are
\begin{align}
    \notag
    \Sigma_{ab}(t) &\equiv \left<N_a(t) N_b(t)\right> - \left<N_a(t)\right> \left<N_b(t)\right> \\
        &= \left(1 - e^{-2r(t-t_f)/N}\right) \left<N_a\right> \left(\delta_{ab} N - \left<N_b\right>\right) + e^{-2r(t-t_f)/N} \Sigma_{ab}(t_f)
         \qquad t \ge t_f.
\end{align}
Expanding this to largest order in $N$, yields
\begin{align}
    \Sigma_{ab}(t) \sim 2 r \left(t-t_f\right) \left(\delta_{ab} \left<N_a\right> - \frac{1}{N} \left<N_a\right> \left<N_b\right>\right) + \Sigma_{ab}(t_f)
         \qquad t \ge t_f.
\end{align}
Note that the expansion in $N$ converges only for finite times.

\newpage
\bibliographystyle{unsrt}
\bibliography{coll.bib}

\end{document}